\documentclass[12pt]{article}
\usepackage{graphicx}
\usepackage{mathrsfs}
\usepackage{dsfont}
\usepackage{graphicx}
\usepackage{units}
\usepackage{amsmath, amsthm, amssymb, amsfonts, enumerate}
\usepackage{natbib}
\usepackage{color}
\usepackage{caption}
\usepackage{subcaption}
\usepackage{setspace}
\usepackage{mdframed}
\usepackage{hyperref}
\usepackage{float}
\usepackage{multirow} 
\usepackage[normalem]{ulem}
\usepackage{placeins}
\usepackage{tikz}
\usetikzlibrary{calc}

\definecolor{blue}{rgb}{0.0, 0.0, 1}
\definecolor{green2}{rgb}{0.0, 0.52, 0.24}
\definecolor{cadmiumgreen}{rgb}{0.0, 0.42, 0.24}
\definecolor{camouflagegreen}{rgb}{0.47, 0.53, 0.42}
\definecolor{darkolivegreen}{rgb}{0.33, 0.42, 0.18}
\definecolor{darkpastelgreen}{rgb}{0.01, 0.75, 0.24}
\definecolor{darkspringgreen}{rgb}{0.09, 0.45, 0.27}
\definecolor{darkspringgreen}{rgb}{0.09, 0.45, 0.27}

\usepackage{comment} 
\usepackage[textwidth=30mm,disable]{todonotes}

\newcommand{\argmax}{{\rm argmax}}

\def\o{{\omega}}

\def\LB{\bigl\{ }
\def\RB{\bigr\} }

\newcommand{\heading}[1]{\vskip .3cm \noindent \textbf {#1} \hskip .4cm }

\def\argmax{\rm{argmax}}

\long\def\ignore#1{}
\def\a{\alpha}
\def\b{\beta}
\def\l{\ell}

\def\?{\Red{?????????}}
\def\1{\textbf{1}}

\newtheorem{theorem}{Theorem}

\newtheorem{proposition}{Proposition}

\newtheorem{example}{Example}
\newtheorem{definition}{Definition}
\newtheorem{claim}{Claim}

\newtheorem{lemma}{Lemma}
\newtheorem{obs}{Observation}

\begin{document}

\title{Comparison of Oracles: Part II\thanks{For their valuable comments, the authors wish to thank participants of the Durham University Economics Seminar, the Adam Smith Business School Micro theory seminar of Glasgow University, INSEAD EPS seminar, the Tel-Aviv University Game Theory Seminar, the Rationality Center Game Theory Seminar, the Technion Game Theory Seminar, the Bar-Ilan University Theoretical Economics Seminar, the Bar-Ilan University Management Seminar, and the BGU Economics seminar.
Lagziel acknowledges the support of the Israel Science Foundation, Grant \#2074/23. Lehrer acknowledges the Israel Science Foundation, Grant \#591/21. Wang acknowledges the support of the National Natural Science Foundation of China \#72303161.}}
\author{David Lagziel\thanks{Department of Economics, Ben-Gurion University of the Negev, Beer-Sheba 8410501, Israel.  E-mail: \textsf{Davidlag@bgu.ac.il}.} \\
{\small Ben-Gurion University}
\and
Ehud Lehrer\thanks{Economics Department, Durham University, Durham DH1 3LB, UK.  E-mail: \textsf{ehud.m.lehrer@durham.ac.uk}.} \\ {\small Durham University}
\and
Tao Wang\thanks{International School of Economics and Management, Capital University of Economics and Business, Beijing 100070, China.  E-mail: \textsf{tao.wang.nau@hotmail.com}.}  \\
{\small CUEB}}
\maketitle

\thispagestyle{empty}

\begin{abstract}
\singlespacing{
This paper studies incomplete-information games in which an information provider, an oracle, publicly discloses information to the players. 
One oracle is said to dominate another if, in every game, it can replicate the equilibrium outcomes induced by the latter. 
The companion Part I characterizes dominance under deterministic signaling and under stochastic signaling with a unique common knowledge component.
The present paper extends the analysis to general environments and provides a characterization of equivalence (mutual dominance) among oracles.
To this end, we develop a theory of information loops, thereby extending the seminal work of \cite{Blackwell1951} to strategic environments and \cite{Aumann1976}'s theory of common knowledge.}
\end{abstract}

\bigskip
\noindent {\emph{Journal of Economic Literature} classification numbers: C72, D82, D83.}

\bigskip

\noindent Keywords:  oracle; information dominance;  signaling function; common knowledge component, information loops.

\newpage
\setcounter{page}{1}


\section{Introduction} \label{Section - Intro}


In settings with incomplete information, whether in peace negotiations, business decisions, or financial markets, players lack full knowledge of all factors that influence the outcomes of their decisions. 
To address such environments, specialized information providers (e.g., peace mediators, business consultants, and rating agencies) operate as neutral \emph{oracles}, selectively disclosing relevant information that can alter strategic behavior and equilibrium outcomes. 
This paper studies the role of such oracles in games of incomplete information, modeling them as agents who transmit information through general signaling functions to the players.

Our primary objective is to characterize when one oracle \emph{dominates} another and when two oracles are \emph{equivalent}.
To this end, we define a partial order of dominance: one oracle dominates another if, in every game, the information structure of the former can induce the same set of equilibrium outcomes as the latter.
Naturally, oracles are equivalent under mutual dominance.\footnote{Note that we abstract away from cases in which the oracle has preferences over players’ action profiles or derives utility from their strategic interaction. In this sense, we adopt Blackwell’s approach (see~\citealp{Blackwell1951}), which focuses on comparing signaling structures (namely, experiments) in decision problems, independently of the sender’s objectives.}

Building on~\cite{Aumann1976}, the notion of a \emph{common knowledge component} (CKC), i.e., the inclusion-wise smallest set that all players can agree upon, plays a central role in our analysis.
The intuition for this is rather clear. 
In an incomplete information game, the relevant set of states for strategic consideration is the corresponding CKC, however an oracle's knowledge is not confined to it.
Oracles, who typically possess information that the players do not, cannot always distinguish between states located in different CKCs. 
Thus, the structure of CKCs governs the interplay between the players’ subjective knowledge and the oracle’s informational limitations.

The CKC also defines the boundary between the companion Part~I (i.e., \citealp{Lagziel2025d}) and the present paper.\footnote{Throughout the paper, we sometimes refer to \cite{Lagziel2025d} as ``Part I".}
Specifically, Part~I characterizes dominance when oracles are restricted to deterministic signaling functions, and when stochastic signals are permitted but the state space features a unique CKC. 
Here, we extend the analysis to environments with multiple CKCs and, in addition, provide a general characterization of equivalence.

Using the structure of multiple CKCs, we introduce the concept of an information loop, the second key element in our characterization.
To formally define these loops and present the main results of the current study, we first partition the state space into distinct CKCs.
An information loop is then defined as a closed path of states that connects different CKCs through elements of an oracle's partition.

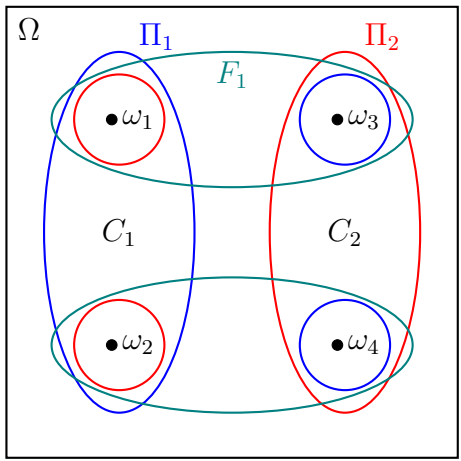
\begin{figure}[th!]
\centering
\centering
\begin{tikzpicture}[scale=1]

\draw[thick] (0,0) rectangle (6,6);

\node at (0.3,5.7) {$\Omega$};

\node[blue] at (2,5.6) {$\Pi_1$};

\node[red] at (5,5.6) {$\Pi_2$};

\draw[blue, thick] (1.5,3) ellipse (1 and 2.4);

\draw[red, thick] (1.5,4.5) ellipse (0.6 and 0.6);

\draw[red, thick] (1.5,1.5) ellipse (0.6 and 0.6);

\draw[red, thick] (4.5,3) ellipse (1 and 2.4);

\draw[blue, thick] (4.5,4.5) ellipse (0.6 and 0.6);

\draw[blue, thick] (4.5,1.5) ellipse (0.6 and 0.6);

\node[teal] at (3,5.1) {$F_1$};

\node[black] at (1.5,3) {$C_1$};
\node[black] at (4.5,3) {$C_2$};

\draw[teal, thick] (3,4.5) ellipse (2.4 and 0.9);
\draw[teal, thick] (3,1.5) ellipse (2.4 and 0.9);

\filldraw[black] (1.4,4.5) circle (2pt) node[anchor=west] {$\o_1$};
\filldraw[black] (1.4,1.5) circle (2pt) node[anchor=west] {$\o_2$};
\filldraw[black] (4.4,4.5) circle (2pt) node[anchor=west] {$\o_3$};
\filldraw[black] (4.4,1.5) circle (2pt) node[anchor=west] {$\o_4$};

\end{tikzpicture}
\caption{There are two CKCs $\{\omega_1,\omega_2\}$ and $\{\omega_3,\omega_4\}$. The oracle's partition $F_1$ generates a loop $(\omega_1,\omega_3,\omega_4,\omega_2)$, which is a closed path connecting the two CKCs using the oracle's partition elements.}
\label{fig: an information loop for the introduction}
\end{figure}

For example, assume there are $4$ states $\Omega =\{\omega_1,\omega_2,\omega_3,\omega_4\}$ and two players whose private information is given by the following partitions: $\Pi_1=\{\{\omega_1, \omega_2\},\{\omega_3\},\{\omega_4\}\}$ and $\Pi_2=\{\{\omega_1\},\{\omega_2\},\{\omega_3,\omega_4\}\}$.
The players' private information induces two CKCs: $C_1 = \{\omega_1,\omega_2\}$ and $C_2=\{\omega_3,\omega_4\}$.
That is, the two players can agree on each of these two events.
See the illustration in Figure \ref{fig: an information loop for the introduction}.
If the oracle's information is given by the partition $F_1= \{\{\omega_1,\omega_3\},\{\omega_2,\omega_4\}\}$, we say that a loop exists, as the different partition elements of $F_1$ form a closed path between the two CKCs.
Namely, $\o_1\in C_1$ and $\o_3 \in C_2$ are joined by a partition element of $F_1$ and the same holds for $\o_2 \in C_1$ and $\o_4 \in C_2$.
This yields a sequence of states that starts in $C_1$, transitions to $C_2$, and reverts back again to $C_1$, through different states that serve as entry and exit points from each CKC.

Assuming that an oracle does not generate information loops, which includes the case where the entire state space comprises a unique CKC, we prove that it dominates the other oracle if and only if its partition refines that of the other within every CKC (see Theorem \ref{Theorem - no loop characterization} in Section \ref{Section - Multiple common knowledge components and no loops}).
Importantly, this result extends the characterization result of Part~I given a unique CKC, while the refinement condition does not follow from the criterion used in the deterministic setting.

\ignore{At this stage, we also prove that the refinement and dominance notions, given a unique CKC, are both equivalent to the \emph{inclusion condition} which states that for every signaling strategy $\tau_2$ of Oracle $2$, there exists a signaling function $\tau_1$ of Oracle $1$ such that the set of the players' posterior beliefs profiles based on $\tau_1$ is a subset of that based on $\tau_2$.
Again, this holds given a single CKC, which obviously cannot admit information loops.}

However, if a loop exists, the characterization becomes more complex.
An information loop imposes (measurability) constraints on the information the oracle can convey.
In the previous example, notice that every signaling function of the oracle over $\{ \o_1,\o_2\}$ uniquely defines the signaling over $\{\o_3,\o_4\}$.
Thus, the oracle is not free to signal any information it wants in one CKC, without restricting its ability to convey different information in the other CKC.

An obvious question that goes to the heart of information loops and our results is, why should we care specifically about the signaling structure over the \emph{pairs of states} that form the loop in every CKC? Moreover, why should a loop consist of separate entry and exit points in every CKC?
The answer is that, given a CKC, Bayesian updating depends on the ratio of signal-probabilities for the different states.
Thus, an effective constraint imposes restrictions over such ratios, thus relating to at least two states in every CKC (while keeping in mind the refinement condition in every CKC; this is a crucial aspect in \citealp{Lagziel2025f}).

The concept of information loops hints at a significant connection to Aumann’s theory of common knowledge, from \cite{Aumann1976}.
This link appears to be central to understanding how shared and differing information structures impact equilibrium outcomes in incomplete-information games.
For this reason we provide an extensive set of results concerning various properties of information loops in Section \ref{Section - Information Loops}.

Specifically, the first property of information loops that we study is \emph{non-informativeness}.
A loop is called \emph{non-informative} if, in every CKC that it intersects, all the states of the loop are in the same partition element of that oracle.
We refer to this as non-informativeness because, conditional on the CKC and loop, the oracle has no information to convey to the players.
For example, in Figure \ref{fig: an information loop for the introduction}, consider an oracle with a trivial partition $F_1'=\Omega=\{\o_1,\o_2, \o_3, \o_4\}$.
This partition creates a closed path between the two CKCs, as well as joining all the states of the loop (given a CKC) to a single partition element of $F_1'$.
Building on this notion and assuming that the partition of Oracle $1$ refines that of Oracle $2$ in every CKC, as in the previously stated characterization, then non-informative loops do not pose a problem for dominance and Oracle $1$ dominates the other (see Theorem \ref{Theorem: NI leads to dominance} in Section \ref{Section - necessary and sufficient conditions for dominance}).

\begin{figure}
        \centering
        \begin{tikzpicture}[scale=0.8]

        \draw (1.5,0.5) -- (4.5,0.5) -- (4.5,1.5) -- (1.5,1.5) -- cycle;
        \node at (2,1) {$\o_1$};
        \node at (4,1) {${\o}_2$};

        \draw (4.5,-1) -- (5.5,-1) -- (5.5,-3.5) -- (4.5,-3.5) -- cycle;
        \node at (5,-1.5) {$\o_3$};
        \node at (5,-3) {$\o_4$};

        \draw (0.5,-1) -- (1.5,-1) -- (1.5,-3.5) -- (0.5,-3.5) -- cycle;
        \node at (1,-3) {$\o_5$};
        \node at (1,-1.5) {$\o_6$};

        \draw[blue, thick, rotate around={-65:(4.45,-0.25)}] (4.45,-0.25) ellipse (2.0cm and 0.7cm); 
        \node[blue] at (4.4,-0.2) {$F_1(\omega_2)$};
        \draw[red, thick, rotate around={0:(3,-3)}] (3,-3) ellipse (2.4cm and 0.7cm); 
        \node[red] at (3,-3) {$F_1(\omega_4)$};
        \draw[teal, thick, rotate around={65:(1.5,-0.25)}] (1.5,-0.25) ellipse (2.0cm and 0.7cm); 
        \node[teal] at (1.5,-0.2) {$F_1(\omega_4)$};

        \node at (3,1.8) {$C_1$};
        \node at (6,-2.5) {$C_2$};
        \node at (-0.1,-2.5) {$C_3$};

        \end{tikzpicture}

     \caption{ \footnotesize An illustration of a fully-informative and irreducible loop, which intersects three CKCs $C_1, C_2$ and $C_3$ with two states in each.}
    \label{fig: F1 irreduicble loop}
\end{figure}
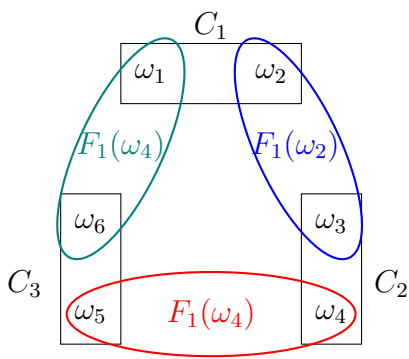

However, once a loop is \emph{informative} (i.e., in at least one CKC that it intersects, there are states in the loop from different partition elements of the oracle; see Figure \ref{fig: F1 irreduicble loop}), then we require additional conditions for characterization.
More specifically, in case there are only two CKCs, an additional condition is that Oracle $2$ also has information loops whose states cover Oracle $1$'s loop, roughly stating the up to non-informative set of states, Oracle $2$ has similar loops to those of Oracle $1$ (the notion of a cover is formally defined in Section \ref{Section - Information Loops}).
Using this condition we provide a characterization for the case of two CKCs (see Proposition \ref{Proposition: two CKCs} in Section \ref{Section - Toward a general characterization: two CKCs}). 
While the question of characterization in the case of more than two CKCs remains open, we do provide necessary conditions for dominance in the general case in Theorem \ref{Theorem: general case -- necessary condition}, building on the notion of \emph{irreducibility}.

The notion of irreducibility, which proves crucial for our analysis, splits to two levels.
The first is \emph{irreducible loops}, which implies that there exists no (smaller) loop that is based on a strict subset of states taken from the original loop.
The second is referred to as \emph{type-2 irreducible loops}, and it implies that the loop does not contain four states from the same partition element of the oracle (again see Figure \ref{fig: F1 irreduicble loop}).
On the one hand, type-2 irreducibility is a weaker notion compared to irreducible loops, because it allows for a loop to intersect the same CKC several times, whereas an irreducible loop cannot.
On the other hand, a type-2 irreducible loop must be informative because it does not allow for the entry and exit point in every CKC to be in the same partition element of that oracle.
In fact, it is \emph{fully-informative} because this condition holds in every CKC, rather than in a specific CKC.

The somewhat delicate understanding of the relations between these loops properties allows us to achieve another main result: the characterization of equivalent oracles.
Formally, we say that two oracles are equivalent if they simultaneously dominate one another.
The characterization of equivalence, given in Theorem \ref{Theorem: Equivalent oracles} in Section \ref{Section - Equivalent oracles}, is based on: (i) equivalence in every CKC; (ii) equivalence of irreducible-informative loops; and (iii) a cover over loops.
To prove this result, we use type-2 irreducible loops to compare the information of both oracles.
Specifically, we consider the sets of type-2 irreducible loops that intersect a joint CKC (i.e., \emph{connected loops}), also taking into account sequential intersections (i.e., the transitive closure) where loop $1$ is connected to loop $2$ which is then connected to loop $3$ and so on.
We observe the set of CKCs for each of these groups and refer to these sets as \emph{clusters}.
These are used as building blocks in our analysis, and we prove that the information of equivalent oracles must match on these clusters.
This, in turn, provides some insight into the possible future characterization of general dominance between oracles, as well as provides another level of extending the theory of common knowledge, beyond information loops.

\subsection{Relation to literature}

Part II takes the comparison of oracles beyond the two benchmark environments handled in Part I (that is, beyond deterministic signaling and stochastic signaling on a state space with a single CKC), and develops tools for general stochastic signaling when multiple CKCs interact. 
The central contribution is the introduction of information loops and associated notions: balance, covers, irreducibility (including type-2 irreducibility), and cluster-based aggregation, which together deliver necessary and sufficient conditions in the presence of loops, and a full equivalence characterization that builds on order-preserving covers of irreducible, fully-informative loops. 

Our starting point remains Blackwell’s comparison of experiments (see \citealp{Blackwell1951,Blackwell1953}), but the object of comparison and the criterion differ in two key ways.
First, an oracle is an \emph{experiment generator}, namely, it can implement any public experiment measurable with respect to its partition, rather than being a fixed experiment. 
Second, the criterion is \emph{strategic and multi-player}, so dominance is defined by equality of the sets of Nash-equilibrium outcome distributions across all games, holding players' private partitions fixed. 
These differences matter only weakly with a single CKC, but are crucial with multiple CKCs, where the loop calculus captures exactly how measurability forces cross-component co-movement of posteriors.

Our CKC-based analysis traces back to the epistemic foundations of games, interacting specifically with the common knowledge ideas of \cite{Aumann1976}. 
For Part~II, where the state space decomposes into multiple CKCs, the right lens is the approximation of common knowledge by common beliefs \emph{\`a la} \citet{Monderer1989}, who formalize $p$-belief and common $p$-belief, showing how implications that classically require exact common knowledge can be approximated by sufficiently strong common beliefs.
The work of Aumann was also followed by \citet{Mertens1985}, who construct a universal type space embedding all coherent hierarchies of beliefs, thus providing a unified measurable framework for Bayesian games, and by \citet{Brandenburger1993}, who clarify the equivalence between hierarchies of beliefs and type representations, linking them to common knowledge. 

Our model builds on these studies by fixing the partition structures while varying only the oracle's public experiment. 
The novel constraints we study arise from \emph{global} measurability across CKCs (via loops), not from additional complexity in private belief hierarchies.
Our information loops formalize when public measurability (via the oracle's information) stitches distinct CKCs so that posterior ratios must align across components, and when such stitching is slack (no loops) or binding (informative, irreducible loops). 
This conceptual bridge clarifies why refinement within CKCs suffices absent loops, but not in general.

Relative to information design and persuasion, the present analysis is comparative rather than optimal. 
The persuasion literature\footnote{For a recent survey, see \cite{Kamenica2019}.} asks which experiment maximizes a sender's objective. 
Here the oracle has no objective, but is evaluated by its replication ability.
In this sense, this project complements persuasion by characterizing when two generators of public experiments are equivalent or when one dominates another.

Closer to us, \citet{Kolotilin2017} analyze persuasion with a privately informed receiver and establish conditions under which optimal mechanisms can be represented as experiments, delivering tractable characterizations in linear/monotone environments. 
Part~II treats the players' experiments as primitives, but evaluates an oracle by the ability to replicate another across all games with fixed private information, so that the binding obstacles are global measurability (loops) rather than incentive constraints.

Another strand in the literature studies mediators in games with incomplete information. 
Mediators deliver differential recommendations that coordinate players’ actions and implement variants of correlated equilibria \citep{Forges1993}. 
In many formulations the mediator does not convey additional information about the realized state; i.e., its role is purely coordinative. 
Under complete information, \citet{Gossner2000} compares mediating structures by the sets of correlated equilibria they can induce, calling one device “richer’’ if it generates a superset. 
This characterization uses a notion of compatible interpretation in the spirit of garbling. 
Part~II departs from this strand in two respects: the oracle’s messages are public and informational about the state, and comparison is by replication power across all games. 
With multiple CKCs, feasibility is governed not by recommendation schemes but by measurability links across CKCs, captured in our framework by information loops (balance and covers).

Closer to the present project are studies on incomplete-information games that establish partial orderings of information structures. 
\citet{Peski2008} obtains a Blackwell-type ordering in zero-sum games. 
\citet{Lehrer2010} analyze common-interest games with privately observed, possibly correlated signals, showing that comparative results hinge on the version of Blackwell garbling tied to the chosen solution concept. 
\citet{Lehrer2013} extend garbling to characterize \emph{outcome equivalence}. 
\citet{Bergemann2016} study $n$-player environments via Bayes correlated equilibrium and characterize dominance through \emph{individual sufficiency}. 
Part~II differs along two margins crucial with multiple CKCs: (i) players’ private partitions are fixed primitives while the oracle is an experiment generator of public signals; and (ii) dominance/equivalence are defined by the ability to reproduce the set of equilibrium outcome distributions in \emph{every} game, and thus hinge on the loop calculus rather than garbling alone.

\heading{The structure of the paper.}
The paper is organized as follows.
Section \ref{Section - Model} depicts the model.
Section \ref{Section - Multiple common knowledge components and no loops} provides a characterization of dominance when there are no loops.
Section \ref{Section - Information Loops} studies the properties of information loops.
Section \ref{Section - Information Loops} outlines necessary and sufficient conditions for dominance, as well as a characterization of dominance given two CKCs (in Section \ref{Section - Toward a general characterization: two CKCs}).
Finally, in Section \ref{Section - Equivalent oracles} we characterize the equivalence relation between oracles.
Appendix \ref{Appendix_Part I results} reviews several key results from the companion Part I. The remainder of the appendix contains the proofs.

\section{The model} \label{Section - Model}


A \emph{guided game} consists of a Bayesian game together with an \emph{oracle}.
The oracle provides information intended to enable a different, and preferably broader, set of equilibria.
It operates via signaling, and our analysis characterizes the extent to which oracles can expand the set of equilibrium payoffs.


We begin by defining the underlying Bayesian game.
Let $N=\{1,2,\dots,n\}$ be a finite set of $n\ge 2$ players, and let $\Omega$ be a non-empty, finite state space.
Each player $i\in N$ has a non-empty, finite action set\footnote{In this framework, $A_i$ is independent of the player's information, but the setting can also accommodate cases where it is not.} $A_i$ and an information partition $\Pi_i$ of $\Omega$.
Let $A=\times_{i\in N} A_i$ denote the set of action profiles.
Player $i$'s utility is $u_i:\Omega\times A\to\mathbb{R}$, mapping states and action profiles to payoffs.


To extend the basic game to a guided game, we introduce an oracle that provides public information before actions are chosen.
The oracle has a partition $F$ of $\Omega$ and a countable signal set $S$.
A signaling strategy of the oracle is an $F$-measurable function $\tau:F\to\Delta(S)$ with finite-support distributions, used to transmit information to all players $N$, where $\Delta(S)$ denotes the set of finite-support probability distributions over $S$.
For $\omega\in\Omega$ and $s\in S$, we write $\tau(s\mid \omega)$ for the probability $\tau(\omega)(s)$ that $s$ is sent when the realized state is $\omega$.
Note that any deterministic strategy $\tau:F\to S$ is effectively a partition, and we refer to it as such when appropriate.


The guided game evolves as follows.
First, the oracle publicly announces a signaling strategy $\tau$.
Then, a state $\omega\in\Omega$ is drawn according to a common prior $\mu\in\Delta(\Omega)$.
Each player $i$ is privately informed of $\Pi_i(\omega)$, the atom (i.e., set of states) of player $i$'s partition that contains $\omega$.
Finally, a realization $s\in S$ is drawn according to $\tau(\omega)$ and publicly announced.

Let the join\footnote{Coarsest common refinement of $\Pi_i$ and $F'$; following \cite{Aumann1976}.} $\Pi_i\vee F'$ denote the updated partition of player $i$ given $\Pi_i$ and a partition $F'$.
If $\tau$ is deterministic, define $\mu^i_{\tau\mid \omega}=\mu\bigl(\cdot\mid [\Pi_i\vee \tau](\omega)\bigr)\in\Delta(\Omega)$ as player $i$'s posterior after observing $\Pi_i(\omega)$ and $\tau(\omega)$.
If $\tau$ is stochastic, let $\mu^i_{\tau\mid \omega,s}=\mu(\cdot\mid \Pi_i(\omega),\tau,s)\in\Delta(\Omega)$ denote player $i$'s posterior after observing $\Pi_i(\omega)$ and a realized signal $s$ according to $\tau(\omega)$.
Thus, every strategy $\tau$ induces an incomplete-information game
$G(\tau)=(N,(A_i)_{i\in N},(\mu^i_{\tau})_{i\in N},(u_i)_{i\in N})$.
Since the state space and action sets are finite, the Nash equilibria exist.
When there is no risk of ambiguity, we denote the incomplete-information game without $\tau$ by $G$.

\subsection{Partial ordering of oracles}

To discuss the oracle's role in this framework, we adopt a solution concept, referred to as a \emph{Guided equilibrium}, that incorporates the oracle's strategy.
Let $\sigma_i:\Pi_i\times S\rightarrow \Delta(A_i)$ be a strategy for player $i$.
A tuple $(\tau,\sigma_1,\dots,\sigma_n)$ is a \emph{Guided equilibrium} if $(\sigma_1,\dots,\sigma_n)$ is a Nash equilibrium of the incomplete-information game $G(\tau)$.

This notion of a Guided equilibrium induces a partial order over oracles (that is, over their partitions) via the sets of equilibria they can generate.
Let $\mathrm{NED}(G(\tau))\subseteq \Delta(\Omega\times A)$ denote the set of distributions over $\Omega\times A$ induced by Nash equilibria given $G$ and $\tau$.\footnote{A Nash equilibrium $(\sigma_1^*,\ldots,\sigma_n^*)$, along with the common prior $\mu$, induce a probability distribution on $\Omega\times A$.
Fix $\omega$ and an action profile $a$.
The probability of $(\omega,a)$ under $(\sigma_1^*,\ldots,\sigma_n^*)$, $\tau$ and the common prior $\mu$ equals $\mu(\omega)\sum_{s\in S}\tau(s\mid \omega)\prod_{i=1}^n \sigma_i^*(a_i\mid \Pi_i(\omega),s)$.
As multiple equilibria may exist, $\mathrm{NED}(G(\tau))$ is a subset of $\Delta(\Omega\times A)$.}
Now consider two oracles, Oracle $1$ and Oracle $2$, and let $F_j$ and $\tau_j$ denote the partition and strategy of Oracle $j$, respectively.
Using these notations, we define a partial order as follows.

\begin{definition}[Partial ordering of Oracles] \label{Definition - Strategic Dominance}
    \emph{Oracle $1$ dominates Oracle $2$}, denoted $F_1 \succeq_{\rm{NE}} F_2$, if for every $\tau_2$ and game $G$, there exists $\tau_1$ such that $\rm{NED}(G(\tau_1)) = \rm{NED}(G(\tau_2))$.
\end{definition}

Informally, dominance means that one oracle can replicate the other’s signaling structure so as to induce the same set of equilibrium outcomes.
A direct comparison of equilibria across games without conditioning on the signaling rule is problematic because players’ strategies typically depend on the oracle’s signals.

\subsection{More than one CKC: two examples}

The partition-refinement condition given in \cite{Lagziel2025d} ensures that Oracle $1$ can produce the \emph{exact} same strategy as Oracle $2$.
This however, hinges on the existence of a unique CKC.
In case there are several CKCs, Oracle $1$ may need to follow a different strategy in order to match the distribution on posteriors generated by $\tau_2$.
Namely, $\tau_1$ may require more signals than $\tau_2$, even if both oracles have the same (complete) information in every CKC.
Let us provide a concrete example for this.

\begin{example} More signals are needed. \label{ex:More than one CKC} \end{example}

Consider a uniformly distributed state space $\Omega =\{\o_1,\o_2,\o_3,\o_4\}$, with two players whose private information is $\Pi_1=\{\{\o_1,\o_2\},\{\o_3\},\{\o_4\}\}$ and $\Pi_2=\{\{\o_1\},\{\o_2\},\{\o_3,\o_4\}\}$.
The oracles have the following partitions $F_1 = \{\{\o_1,\o_3\}, \{\o_2\},\{\o_4\}\}$ and $F_2 = \{\{\o_1\},\{\o_3\}, \{\o_2,\o_4\}\}$.
This information structure is illustrated in Figure \ref{fig: more than one CKC an example}.
Notice that there are two CKCs, $\{\o_1,\o_2\}$ and $\{\o_3,\o_4\}$, and both oracles have complete information in each of these components.
That is, $F_1$ refines $F_2$ in every CKC, and vice versa.

\begin{figure}[th!]
\centering
\begin{minipage}{0.45\textwidth}
\centering
\begin{tikzpicture}[scale=1]

\draw[thick] (0,0) rectangle (6,6);

\node at (0.3,5.7) {$\Omega$};

\node[blue] at (2,5.6) {$\Pi_1$};

\node[red] at (5,5.6) {$\Pi_2$};

\draw[blue, thick] (1.5,3) ellipse (1 and 2.5);

\draw[red, thick] (1.5,4.5) ellipse (0.6 and 0.8);

\draw[red, thick] (1.5,1.5) ellipse (0.6 and 0.8);

\draw[red, thick] (4.5,3) ellipse (1 and 2.5);

\draw[blue, thick] (4.5,4.5) ellipse (0.6 and 0.8);

\draw[blue, thick] (4.5,1.5) ellipse (0.6 and 0.8);

\filldraw[black] (1.4,4.5) circle (2pt) node[anchor=west] {$\o_1$};
\filldraw[black] (1.4,1.5) circle (2pt) node[anchor=west] {$\o_2$};
\filldraw[black] (4.4,4.5) circle (2pt) node[anchor=west] {$\o_3$};
\filldraw[black] (4.4,1.5) circle (2pt) node[anchor=west] {$\o_4$};

\end{tikzpicture}
\caption*{(a)}
\vspace{-0.2cm}
\caption*{The players' information}
\end{minipage}%
\hfill
\begin{minipage}{0.45\textwidth}
\centering
\begin{tikzpicture}[scale=1]

\draw[thick] (0,0) rectangle (6,6);

\node at (0.3,5.7) {$\Omega$};

\node[orange] at (3,5.1) {$F_1$};

\node[teal] at (3,2.1) {$F_2$};

\draw[orange, thick] (1.5,1.5) ellipse (0.7 and 0.7);
\draw[orange, thick] (4.3,1.5) ellipse (0.7 and 0.7);

\draw[orange, thick] (3,4.5) ellipse (2.7 and 1);

\draw[teal, thick] (3,1.5) ellipse (2.7 and 1);

\draw[teal, thick] (1.5,4.5) ellipse (0.7 and 0.7);
\draw[teal, thick] (4.3,4.5) ellipse (0.7 and 0.7);

\filldraw[black] (1.5,4.5) circle (2pt) node[anchor=west] {$\o_1$};
\filldraw[black] (1.5,1.5) circle (2pt) node[anchor=west] {$\o_2$};
\filldraw[black] (4.3,4.5) circle (2pt) node[anchor=west] {$\o_3$};
\filldraw[black] (4.3,1.5) circle (2pt) node[anchor=west] {$\o_4$};

\end{tikzpicture}
\caption*{(b)}
\vspace{-0.2cm}
\caption*{The oracles' information}
\end{minipage}
\caption{On the left, Figure (a) illustrates the information structure of player $1$ (blue) and player $2$ (red). On the right, Figure (b) portrays the information structure of Oracle $1$ (orange) and Oracle $2$ (green).}
\label{fig: more than one CKC an example}
\end{figure}

Consider the stochastic strategy $\tau_2$ given in Figure \ref{fig:Table of tau2 in example 2}.
Notice it is $F_2$-measurable, as $\tau_2(s|\o_2)=\tau_2(s|\o_4)$ for every signal $s$, but not $F_1$-measurable.
\begin{figure}[th!]
\centering

\medskip

\begin{tabular}{c|c|c|c|}
    $\tau_2(s|\o)$ & $s_1$ & $s_2$ & $s_3$ \\
\hline
$\o_1$ & 0 & 1/2 & 1/2 \\
\hline
$\o_2$ & 1/3 & 2/3 & 0 \\
\hline
$\o_3$ & 0 & 2/3 & 1/3 \\
\hline
$\o_4$ & 1/3 & 2/3 & 0 \\
\hline
\end{tabular}
\caption{ \footnotesize A stochastic $F_2$-measurable strategy of Oracle $2$.}
\label{fig:Table of tau2 in example 2}
\end{figure}

The set of $\tau_2$-posteriors ${\rm Post}(\tau_2)$ is
\[
{\rm Post}(\tau_2) =
\left\{\!\begin{aligned}
&(e_i, e_i), && \forall \ 1 \leq i \leq 4, \\[1ex]
&\left(\left(\tfrac{3}{7}, \tfrac{4}{7}, 0, 0\right), e_j\right), && j = 1, 2, \\[1ex]
&\left(e_k, (0, 0, \tfrac{1}{2}, \tfrac{1}{2})\right), && k = 3, 4
\end{aligned}\right\},
\]
and we can now try to mimic $\tau_2$ using an $F_1$-measurable strategy.
First, this requires at least two signals to distinguish between $\o_1$ and $\o_2$, as well as $\o_3$ and $\o_4$.
Second, the posterior $\left(\left(\tfrac{3}{7}, \tfrac{4}{7}, 0, 0\right), e_1\right)$ requires another signal $s$ so that $\tau(s|\o_1)=\alpha >0$ and $\tau(s|\o_3)=\tfrac{4}{3}\alpha >0$.
However, the $F_1$-measurability requirement implies that $\tau(s|\o_3)=\alpha$, and the $\tau_2$-posterior $\left(e_3, (0, 0, \tfrac{1}{2}, \tfrac{1}{2})\right)$ necessitates that $\tau(s|\o_4)=\alpha$ as well.
These conditions are jointly given in Table $(a)$ within Figure \ref{fig:Table of tau1 in example 2}.
\begin{figure}[th!]
\centering
\begin{minipage}{0.45\textwidth}
\centering
\begin{tabular}{c|c|c|c|}
    $\tau_1(s|\o)$ & $s_3$ & $s_4$ & $s_5$ \\
\hline
$\o_1$ & $\alpha$ & $\beta$ & $0$ \\
\hline
$\o_2$ & $\tfrac{4}{3} \alpha$  & $0$ & $\gamma$ \\
\hline
$\o_3$ & $\alpha$ & $\beta$ &  $0$ \\
\hline
$\o_4$ & $\alpha$ & $0$ & $\gamma$ \\
\hline
\end{tabular}
\caption*{ \footnotesize (a)}
\end{minipage}%
\hspace{-0.5cm}
\begin{minipage}{0.45\textwidth}
\centering
\begin{tabular}{c|c|c|c|c|}
    $\tau_1(s|\o)$ & $s_3$ & $s_4$ & $s_5$ & $s_6$ \\
\hline
$\o_1$ & $1/2$ & 1/3 & $0$   & $1/6$   \\
\hline
$\o_2$ & $2/3$  & $0$ & $1/3$ & $0$     \\
\hline
$\o_3$ & $1/2$  & $1/3$ &  $0$ & $1/6$  \\
\hline
$\o_4$ & $1/2$ & $0$ & $1/3$ & $1/6$   \\
\hline
\end{tabular}
\caption*{ \footnotesize (b)}
\end{minipage}%
\caption{ \footnotesize A strategy $\tau_1$, either with $3$ signals as given in Table (a), or with $4$ signals as in Table (b).}
\label{fig:Table of tau1 in example 2}
\end{figure}

Evidently, it must be that $\alpha,\beta,\gamma >0$ in order to mimic $\tau_2$, but the second and fourth rows in Table (a) cannot jointly sum to $1$ unless $\alpha=0$, which eliminates the possibility of a well-defined mimicking strategy.
Thus, in order to mimic the stated strategy $\tau_2$, Oracle $1$ requires an additional signal as presented in Table $(b)$, in Figure \ref{fig:Table of tau1 in example 2}.
To conclude, though the oracles' partitions refine one another in every CKC, they cannot always produce the exact same strategy when trying to mimic each other.

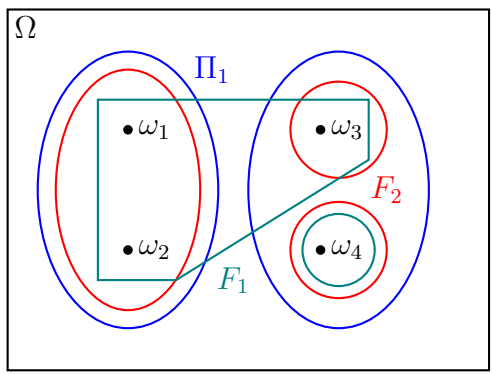
\begin{figure}[th!]
\centering


\medskip

\begin{tikzpicture}[scale=0.8] 

\draw[thick] (0,0) rectangle (8,6);

\node at (0.3,5.7) {$\Omega$};

\node[blue] at (3.4,5) {$\Pi_1$};

\node[red] at (6.3,3) {$F_2$};

\node[teal] at (3.75,1.5) {$F_1$};

\draw[blue, thick] (2,3) ellipse (1.5 and 2.3);

\draw[red, thick] (2,3) ellipse (1.2 and 2);

\draw[red, thick] (5.5,4) ellipse (0.8 and 0.8);
\draw[red, thick] (5.5,2) ellipse (0.8 and 0.8);
\draw[teal, thick] (5.5,2) ellipse (0.6 and 0.6);

\draw[blue, thick] (5.5,3) ellipse (1.5 and 2.3);

\filldraw[black] (2,4) circle (2pt) node[anchor=west] {$\o_1$};
\filldraw[black] (2,2) circle (2pt) node[anchor=west] {$\o_2$};
\filldraw[black] (5.2,4) circle (2pt) node[anchor=west] {$\o_3$};
\filldraw[black] (5.2,2) circle (2pt) node[anchor=west] {$\o_4$};

\draw[teal, thick] (1.5,4.5) -- (6,4.5) -- (6,3.5) -- (2.8,1.5) -- (1.5,1.5) -- cycle;

\end{tikzpicture}
\caption{\footnotesize Note that $F_2$ strictly refines $F_1$ and $\Pi_1$.}
\label{fig:IMI is not finer than}
\end{figure}

\begin{example} Dominance need not imply refinement with multiple \emph{CKCs}
   \label{ex: Stochastic more informative but does not refine}
\end{example}
In this example we wish to show that when there are multiple CKCs, Oracle $1$ can dominate Oracle $2$ although $F_1$ does not refine $F_2$.
To see this, we revisit an example from \cite{Lagziel2025d} in which $ \Pi_1 = \{\{\o_1,\o_2\},\{\o_3,\o_4\}\}, F_1 = \{\{\o_1,\o_2,\o_3\},\{\o_4\}\}$ and $F_2  = \{\{\o_1,\o_2\},\{\o_3\},\{\o_4\}\}$.
This is illustrated in Figure \ref{fig:IMI is not finer than}.

Now consider the  signaling strategy of Oracle $2$ given in Figure \ref{fig:Table of tau2 in example 3}, where Oracle $2$ provides the players with no additional information regarding states $\o_1$ and $\o_2$.
Thus, the posterior over these states remains the original one. On the other hand, given the states $\o_3$ and $\o_4$, the strategy $\tau_2$ reveals the true state with a positive probability and induces the posterior $(0,0, 2/5,3/5)$ with the remaining probability.

\begin{figure}[th!]
\centering


\medskip

\begin{tabular}{c|c|c|c|}
    $\tau_2(s|\o)$ & $s_1$ & $s_2$ & $s_3$ \\
\hline
$\o_1$ & 1/4 & 0 & 3/4 \\
\hline
$\o_2$ & 1/4 & 0 & 3/4 \\
\hline
$\o_3$ & 0 & 1/2 & 1/2 \\
\hline
$\o_4$ & 1/4 & 0 & 3/4  \\
\hline
\end{tabular}
\caption{ \footnotesize A stochastic $F_2$-measurable strategy of Oracle $2$.}
\label{fig:Table of tau2 in example 3}
\end{figure}

While Oracle $2$ can assign different probabilities to a signal conditioned on $\o_2$ and $\o_3$,  Oracle $1$ cannot. However, there is a signaling strategy for Oracle $1$ that produces the same distribution over the posteriors as $\tau_2$ does.
The following strategy $\tau_1$, given in Figure \ref{fig:Table of tau2 in example 4}, does that.

\begin{figure}[th!]
\centering
\begin{tabular}{c|c|c|c|}
    $\tau_1(s|\o)$ & $s_1$ & $s_2$ & $s_3$ \\
\hline
$\o_1$ & 1/2 & 0 & 1/2 \\
\hline
$\o_2$ & 1/2 & 0 & 1/2 \\
\hline
$\o_3$ & 1/2 & 0 & 1/2 \\
\hline
$\o_4$ & 0 & 1/4 & 3/4  \\
\hline
\end{tabular}
\caption{ \footnotesize A stochastic $F_1$-measurable strategy of Oracle $1$.}
\label{fig:Table of tau2 in example 4}
\end{figure}

In this example, it is straightforward to prove that Oracle $1$ can mimic every strategy $\tau_2$ of Oracle $2$, and we prove this result under more general conditions in Theorem \ref{Theorem - no loop characterization} and Proposition \ref{Proposition: two CKCs}.
Yet, it is clear that $F_1$ is not a refinement of $F_2$ in general, but it is a refinement in every CKC.

\section{Multiple CKCs and no loops} \label{Section - Multiple common knowledge components and no loops}

We now turn to the general setting in which the players' information structures induce any (finite) number of CKCs.
Assume that \( C_1, \dots, C_l \) are mutually exclusive CKCs such that \( \Omega = \bigcup_{j=1}^l C_j \).
A key aspect of our analysis is the presence of measurability constraints, where different CKCs are connected by atoms of the oracles' partitions.
To understand the significance of this, consider a setting where \( F_1 \) does not contain any element intersecting multiple CKCs.
In this case, the characterization result given a unique CKC from Part~I (see Theorem \ref{Theorem - a unique CKC} in Appendix \ref{Subsection - a unique CKC}) applies separately to each CKC, as Oracle $1$ faces no constraints when attempting to mimic some strategy of Oracle $2$.

However, when elements of Oracle $1$'s partition intersect different CKCs, the analysis becomes more complex, because we must account for measurability constraints when attempting to use the same strategy \(\tau_1\) across different CKCs.
Such intersections impose constraints on \(\tau_1\), preventing us from naively applying previous results.

This issue becomes even more complicated when multiple elements of Oracle $1$'s partition intersect different CKCs, forming what we call an (information) \emph{loop}.\footnote{ An (information) loop is different from a loop in graph theory. In graph theory, a loop refers to an edge that connects a vertex to itself.}

Generally, a loop is an ordered sequence of states from different CKCs such that the partition of an oracle groups together distinct pairs of states from different CKCs, creating a closed path.
The main result of this section, presented in Theorem \ref{Theorem - no loop characterization} below, states that in the absence of such loops, Oracle $1$ dominates Oracle $2$ if and only if \( F_1 \) refines \( F_2 \) in every CKC.
The formal definition of a loop is provided in Definition \ref{Definition: a loop}.

\begin{definition} \label{Definition: a loop}
    An $F_i$-loop is a sequence $(\omega_1,\overline{\omega}_1, \omega_2,\overline{\omega} _2, \dots, \omega_m,\overline{\omega}_m)$, where $m+1 \equiv 1$ and $m \geq 2$, such that
    \begin{itemize}
        \item $\omega_j,\overline{\omega}_j \in C_{r_j}$ and $\omega_j\neq \overline{\omega}_j$ for all $j=1,\dots,m$.\footnote{Here $C_{r_j}$ refers to the CKC that contains the $j$-th pair of states $(\omega_j,\overline{\omega}_j)$. }
        \item $\omega_{j+1} \in F_i(\overline{\omega}_j)$ for all $j=1,\dots,m$.
        \item $C_{r_j} \neq C_{r_{j+1}}$ for all $j=1,\dots,m$.
        \item The sets $\{\overline{\omega}_j,\omega_{j+1}\}$ are pairwise disjoint for all $j=1,\dots,m$.
    \end{itemize}
\end{definition}

To understand information loops, one can view the CKCs as the vertices of a graph.
An edge connects two CKCs if there exist $\omega_{j+1}$ and $\overline{\omega}_{j}$
such that they belong to the same $F_i$-partition element (this corresponds to the second requirement).
An information loop then parallels an Eulerian graph,
where there is a walk that includes every edge exactly once (the last requirement in the definition)
and ends back at the initial vertex (hence the requirement $m+1\equiv 1$).
As noted at the beginning of this section,
the key aspect of the general analysis is to consider the case when the oracle partition atoms intersect different CKCs,
so we require that $C_{r_j} \neq C_{r_{j+1}}$ for all $j=1,\dots,m$.

An example of an \( F_1 \)-loop is provided in Figure \ref{fig: F1 loop and non-balanced loop}.(a), which depicts a loop consisting of six states across three CKCs.
Note that a loop can intersect the same CKC multiple times, as long as the sets \( \{\overline{\omega}_j, \omega_{j+1}\} \) remain pairwise disjoint for each \( j \).\footnote{A loop intersects a given CKC once if there is a unique pair of states $(\omega_j,\overline{\omega}_j)$ from the loop that lies in that CKC. }

We use the concept of a loop in our first general characterization, presented in Theorem \ref{Theorem - no loop characterization}.
This theorem builds on the assumption that \( F_1 \) contains no loops and extends the main result of Part I by showing that one oracle dominates another if the former’s partition refines that of the latter in every CKC.
It is important to note that the proof is extensive, as it must account for the measurability constraints of \( \tau_1 \) across all CKCs.

\begin{theorem} \label{Theorem - no loop characterization}
    Assume there is no $F_1$-loop. Then, \emph{Oracle} $1$ dominates \emph{Oracle} $2$ if and only if $F_1$ refines $F_2$ in every \emph{CKC}.
\end{theorem}

The proof of Theorem \ref{Theorem - no loop characterization} builds on the concept of a \emph{sub-strategy}.
A sub-strategy is a signaling function without the requirement that the probabilities sum to 1.
This relaxation allows us to study functions that partially mimic a strategy \(\tau_2\), meaning each posterior is drawn from the set of $\tau_2$-posteriors \({\rm Post}(\tau_2)\) and is induced with a probability that does not exceed the probability with which \(\tau_2\) induces it.
We show that the set of sub-strategies is compact, allowing us to consider an optimal sub-strategy for mimicking \(\tau_2\).
The proof then proceeds by contradiction: if the optimal sub-strategy is not a complete strategy, we can extend it by constructing an additional sub-strategy to complement the optimal one for posteriors that are not fully supported (relative to the probabilities induced by \(\tau_2\)).
This part is rather extensive as it requires some graph theory and several supporting claims given in the proof in the appendix.

\begin{figure}
    \centering
    \begin{minipage}{0.45\linewidth}
        \centering
        \begin{tikzpicture}[scale=0.8]

        \draw (1.5,0.5) -- (4.5,0.5) -- (4.5,1.5) -- (1.5,1.5) -- cycle;
        \node at (2,1) {$\o_1$};
        \node at (4,1) {$\overline{\o}_1$};

        \draw (4.5,-1) -- (5.5,-1) -- (5.5,-3.5) -- (4.5,-3.5) -- cycle;
        \node at (5,-1.5) {$\o_2$};
        \node at (5,-3) {$\overline{\o}_2$};

        \draw (0.5,-1) -- (1.5,-1) -- (1.5,-3.5) -- (0.5,-3.5) -- cycle;
        \node at (1,-3) {$\o_3$};
        \node at (1,-1.5) {$\overline{\o}_3$};

        \draw[blue, thick, rotate around={-65:(4.45,-0.25)}] (4.45,-0.25) ellipse (2.0cm and 0.7cm); 
        \node[blue] at (4.4,-0.2) {$F_1(\omega_2)$};
        \draw[red, thick, rotate around={0:(3,-3)}] (3,-3) ellipse (2.4cm and 0.7cm); 
        \node[red] at (3,-3) {$F_1(\omega_3)$};
        \draw[teal, thick, rotate around={65:(1.5,-0.25)}] (1.5,-0.25) ellipse (2.0cm and 0.7cm); 
        \node[teal] at (1.5,-0.2) {$F_1(\omega_1)$};

        \node at (3,1.8) {$C_1$};
        \node at (6,-2.5) {$C_2$};
        \node at (-0.1,-2.5) {$C_3$};

        \end{tikzpicture}
        \caption*{(a)}
    \end{minipage}    
\begin{minipage}{0.45\linewidth}
    \centering
    \begin{tikzpicture}[scale=0.8]

    \draw (1.5,0.5) -- (4.5,0.5) -- (4.5,1.5) -- (1.5,1.5) -- cycle;
    \node at (2,1) {$\o_1$};
    \node at (4,1) {$\overline{\o}_1$};

    \draw (4.5,-1) -- (5.5,-1) -- (5.5,-3.5) -- (4.5,-3.5) -- cycle;
    \node at (5,-1.5) {$\o_2$};
    \node at (5,-3) {$\overline{\o}_2$};

    \draw (0.5,-1) -- (1.5,-1) -- (1.5,-3.5) -- (0.5,-3.5) -- cycle;
    \node at (1,-3) {$\o_3$};
    \node at (1,-1.5) {$\overline{\o}_3$};

    \draw[red, thick] (1.2,0)
    .. controls (1.5,2) and (2,2) .. (3,1)
    .. controls (5,-1) and (5,-1) .. (5.5,-1.4)
    .. controls (5.7,-1.9) and (6,-2.0) .. (0.8,-3.6)
    .. controls (0.3,-3.7) and (0.9,-2.3) .. (1.2,-2)
    .. controls (1.8,-1.3) and (2.0,-1.1) .. (1.6,-0.9)
    .. controls (1.4,-0.7) and (1.1,-0.5) .. (1.2,0);
    \node[red] at (0,0.5) {$A=F_2(\omega_1)$};

    \draw[blue, thick] (4.8,0)
    .. controls (4.5,2) and (4,2) .. (3,1)
    .. controls (1,-1) and (1,-1) .. (0.5,-1.4)
    .. controls (0.3,-1.9) and (0,-2.0) .. (5.2,-3.6)
    .. controls (5.7,-3.7) and (5.1,-2.3) .. (4.8,-2)
    .. controls (4.2,-1.3) and (4,-1.1) .. (4.4,-0.9)
    .. controls (4.6,-0.7) and (4.9,-0.5) .. (4.8,0);
    \node[blue] at (6,0.5) {$B=F_2(\overline{\omega}_1)$};

    \node at (3,1.8) {$C_1$};
    \node at (6,-2.5) {$C_2$};
    \node at (-0.1,-2.5) {$C_3$};

    \end{tikzpicture}
    \caption*{(b)}
\end{minipage}
     \caption{ \footnotesize  Figure (a) depicts an $F_1$-loop with three CKCs and six states overall. Figure (b) illustrates how the $F_1$-loop, presented in $(a)$,  is non-balanced with respect to $F_2$. Namely, $F_2$ has two elements $A=\{\omega_1,\omega_2,\omega_3\}$, and $B=\{\overline{\omega}_1, \overline{\omega}_2, \overline{\omega}_3\}$ such that the number of transitions from $A$ to $B$ are $3$, while the reverse equals $0$.}
    \label{fig: F1 loop and non-balanced loop}
\end{figure}

\section{Information loops} \label{Section - Information Loops}

Previous sections have examined the problem of oracle dominance in the absence of loops, considering either a unique CKC or multiple CKCs without loops.
However, in order to confront the general question of dominance in the presence of information loops, we need to have a clear understanding of their properties and implications.

Specifically, when an \( F_1 \)-loop exists, it may create challenges for Oracle $1$ in mimicking Oracle $2$, because loops introduce measurability constraints across CKCs.
Although Oracle $1$ can mimic Oracle $2$ within each CKC individually, it may be impossible to do so simultaneously across CKCs if the required combined strategy is not measurable with respect to \( F_1 \).
This suggests that any \( F_1 \)-loop must satisfy certain conditions to ensure that such a strategy is indeed \( F_1 \)-measurable.
The first condition that we study, which turns out to be a necessary condition for dominance, is generally referred to as \emph{$F_2$-balanced}.

The idea starts with an \( F_1 \)-loop.
We examine all states in this loop and determine how they can be covered by two \( F_2 \)-measurable sets.
In other words, the loop is divided into two disjoint sets, each contained in an \( F_2 \)-measurable set, denoted \( A \) and \( B \).
Next, we count the number of transitions along the loop from \( A \) to \( B \), where the entry point into one CKC is through a state in \( A \) and the exit is through a state in \( B \).
We do the same for transitions from \( B \) to \( A \).
An \( F_1 \)-loop is called \emph{\( F_2 \)-balanced} if the number of transitions between \( A \) and \( B \) is equal in both directions.
The formal definition follows.

\begin{definition} \label{Definition - F_2 balanced loops}
    An $F_i$-loop $(\omega_1,\overline{\omega}_1, \omega_2,\overline{\omega} _2, \dots, \omega_m,\overline{\omega}_m)$ is $F_{-i}$-\emph{balanced} if for every $F_{-i}$-measurable partition of the loop's states into two disjoint sets $\{A,B\}$ such that $\cup_j\{\o_j, \overline{\omega}_j\} \subseteq A\cup B$, it follows that:
    \begin{equation}\label{eq: balanced}
    \#(A \to B):=|\{j; \o_j\in A \ {\rm{and}} \ \overline{\omega}_j\in B\}|= |\{j; \o_j\in B \ {\rm{and}} \ \overline{\omega}_j\in A\}| =: \#(B \to A).\end{equation}
\end{definition}

Note that an \( F_1 \)-loop \((\omega_1, \overline{\omega}_1, \omega_2, \overline{\omega}_2, \dots, \omega_m, \overline{\omega}_m)\), where \(\omega_{j} \in F_2(\overline{\omega}_j)\) for all \( j = 1, \dots, m \), is \( F_2 \)-\emph{balanced}.
Figure \ref{fig: F1 loop and non-balanced loop}.(b) shows a partition of the $F_1$-loop in  \ref{fig: F1 loop and non-balanced loop}.(a) into two $F_2$-measurable sets $A$ and $B$. Since  \( \#(A \to B) = 3 \) while \( \#(B \to A) = 0 \), the $F_1$-loop fails to be $F_2$-balanced. 

Why are balanced loops crucial? 
Consider, for example, a non-balanced loop as depicted in Figure \ref{fig: F1 loop and non-balanced loop}, and assume that \(\tau_2(s|\omega) = \tfrac{1}{2} - \frac{1}{4} \mathbf{1}_{\{\omega \in A\}}\) for some signal \( s \in S \).
This imposes a specific \(1:2\) ratio between any two states described in each CKC, so that \(\Pi_i \tfrac{\tau_2(s|\omega_i)}{\tau_2(s|\overline{\omega}_i)} = \tfrac{1}{8}\).
However,  since $\overline{\omega}_{i}$ and $\omega_{i+1}$ belong to the same $F_1$ partition element, the measurability constraints on Oracle $1$ along the loop require that $\tau_1(s|\overline{\omega}_{i})=\tau_1(s|{\omega}_{i+1})$, hence \(\Pi_i \tfrac{\tau_1(s|\omega_i)}{\tau_1(s|\overline{\omega}_i)} = 1\) for any \(s\) in the support of all states.
In other words, Oracle $1$ cannot match the ratio dictated by \(\tau_2\), therefore the key proportionality lemma from Part~I (see Lemma \ref{Lemma - proportional signals with m states} from in Appendix \ref{Subsection - proportional signals with m states}) does not hold in at least one CKC.

If the loop were balanced—say, with \( A = \{\overline{\omega}_1, \omega_2\} \) and \( B = \{\omega_1, \overline{\omega}_2, \omega_3, \overline{\omega}_3\} \)—then the same strategy \(\tau_2\) would yield \(\Pi_i \tfrac{\tau_2(s|\omega_i)}{\tau_2(s|\overline{\omega}_i)} = 1\), as required.
In general, when all loops are balanced, this discrepancy is eliminated for any two such sets \( A \) and \( B \).
The notion of balanced loops is closely related to the following notion of \emph{covered loops}, which implies that an $F_1$-loop can be decomposed to loops of $F_2$.

\begin{definition} \label{Definition - covered loop}
    An $F_i$-loop $(\omega_1,\overline{\omega}_1, \omega_2,\overline{\omega}_2, \dots, \omega_m,\overline{\omega}_m)$ is $F_{-i}$\emph{-covered} if
    \begin{itemize}
        \item The set $\{1,...,m\}$ is partitioned to disjoint sets of indices, $J, I_1,...,I_r$, i.e.,  $\{1,...,m\}=J\cup ( \cup_{t=1}^r I_t)$.
        \item For each $t=1,...,r$, $\Big((\omega_{j},\overline{\omega}_j)\Big)_{j\in I_t}$ is an $F_{-i}$-loop, also referred to as a \emph{sub-loop}.\footnote{The order of the pairs $(\omega_{j},\overline{\omega}_j)$  in the $F_{-i}$-loop does not have to coincide with their order under the $F_i$-loop. For instance, an $F_1$-loop $(\omega_1,\overline{\omega}_1, \omega_2,\overline{\omega}_2,  \omega_3,\overline{\omega}_3)$ might be covered by the following $F_2$-loop $(\omega_1,\overline{\omega}_1,  \omega_3,\overline{\omega}_3, \omega_2,\overline{\omega}_2)$. \label{Footnote - order of subloop}}
        \item $J=\{j; \o_j\in F_{-i}(\overline{\omega}_j)\}$.
    \end{itemize}
    The cover is \emph{order-preserving} if every $F_{-i}$-loop $\Big((\omega_{j},\overline{\omega}_j)\Big)_{j\in I_t}$ in the cover follows the same ordering of pairs as the $F_i$-loop.
\end{definition}

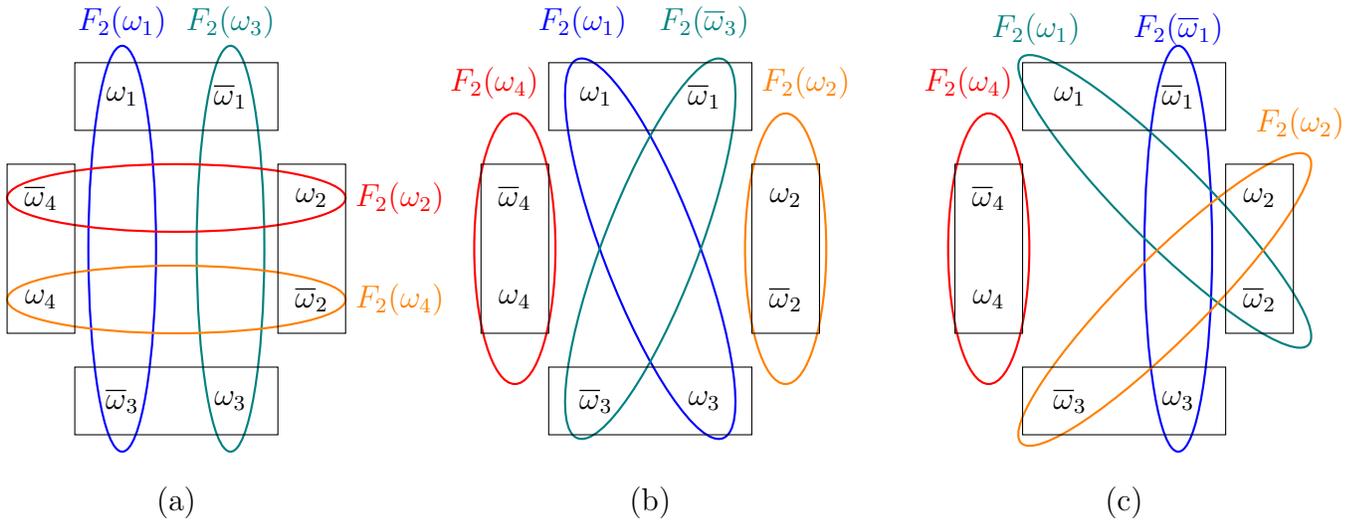
\begin{figure}[ht]
\centering

\begin{tikzpicture}[scale=0.9]

    \draw[blue, thick] (2.2, -1.25) ellipse (0.5 and 3);  
    \draw[teal, thick] (3.8, -1.25) ellipse (0.5 and 3);  

    \draw[red, thick] (3, -0.5) ellipse (2.5 and 0.5);  
    \draw[orange, thick] (3, -2) ellipse (2.5 and 0.5);    

    \draw (1.5,0.5) -- (4.5,0.5) -- (4.5,1.5) -- (1.5,1.5) -- cycle;
    \node at (2.2,1) {$\o_1$};

    \node at (3.8,1) {$\overline{\o}_1$};

    \draw (4.5,0) -- (5.5,0) -- (5.5,-2.5) -- (4.5,-2.5) -- cycle;
    \node at (5,-0.5) {$\o_2$};

    \node at (5,-2) {$\overline{\o}_2$};

    \draw (0.5,0) -- (1.5,0) -- (1.5,-2.5) -- (0.5,-2.5) -- cycle;
    \node at (1,-2) {$\o_4$};

    \node at (1,-0.5) {$\overline{\o}_4$};

    \draw (1.5,-3.0) -- (4.5,-3.0) -- (4.5,-4.0) -- (1.5,-4.0) -- cycle;
    \node at (2.2,-3.5) {$\overline \o_3$};

    \node at (3.8,-3.5) {${\o_3}$};

    \node[blue] at (2.2,2.1) {$F_2(\o_1)$};
    \node[teal] at (3.8,2.1) {$F_2(\o_3)$};
    \node[red] at (6.3,-0.5) {$F_2(\o_2)$};
    \node[orange] at (6.3,-2) {$F_2(\o_4)$};

    \draw[blue, thick, rotate around={-69:(10,-1.25)}] (10,-1.25) ellipse (3.0cm and 0.7cm);
    \draw[teal, thick, rotate around={69:(10,-1.25)}] (10,-1.25) ellipse (3.0cm and 0.7cm);

    \draw[red, thick] (8, -1.25) ellipse (0.6 and 2);  
    \draw[orange, thick] (12, -1.25) ellipse (0.6 and 2);    

    \draw (8.5,0.5) -- (11.5,0.5) -- (11.5,1.5) -- (8.5,1.5) -- cycle;
    \node at (9.2,1) {$\o_1$};

    \node at (10.8,1) {$\overline{\o}_1$};

    \draw (11.5,0) -- (12.5,0) -- (12.5,-2.5) -- (11.5,-2.5) -- cycle;
    \node at (12,-0.5) {$\o_2$};

    \node at (12,-2) {$\overline{\o}_2$};

    \draw (7.5,0) -- (8.5,0) -- (8.5,-2.5) -- (7.5,-2.5) -- cycle;
    \node at (8,-2) {$\o_4$};

    \node at (8,-0.5) {$\overline{\o}_4$};

    \draw (8.5,-3.0) -- (11.5,-3.0) -- (11.5,-4.0) -- (8.5,-4.0) -- cycle;
    \node at (9.2,-3.5) {$\overline \o_3$};

    \node at (10.8,-3.5) {${\o_3}$};

    \node[blue] at (9,2.1) {$F_2(\o_1)$};
    \node[teal] at (10.8,2.1) {$F_2(\overline{\o}_3)$};
    \node[orange] at (12.3,1.2) {$F_2(\o_2)$};
    \node[red] at (7.7,1.2) {$F_2(\o_4)$};

    \draw[blue, thick] (17.8, -1.25) ellipse (0.5 and 3);  

    \draw[teal, thick, rotate around={-45:(17.6,-0.55)}] (17.6,-0.55) ellipse (3.0cm and 0.6cm);
    \draw[orange, thick, rotate around={45:(17.6,-2)}] (17.6,-2) ellipse (3.0cm and 0.6cm);

    \draw[red, thick] (15, -1.25) ellipse (0.6 and 2);  

    \draw (15.5,0.5) -- (18.5,0.5) -- (18.5,1.5) -- (15.5,1.5) -- cycle;
    \node at (16.2,1) {$\o_1$};

    \node at (17.8,1) {$\overline{\o}_1$};

    \draw (18.5,0) -- (19.5,0) -- (19.5,-2.5) -- (18.5,-2.5) -- cycle;
    \node at (19,-0.5) {$\o_2$};

    \node at (19,-2) {$\overline{\o}_2$};

    \draw (14.5,0) -- (15.5,0) -- (15.5,-2.5) -- (14.5,-2.5) -- cycle;
    \node at (15,-2) {$\o_4$};

    \node at (15,-0.5) {$\overline{\o}_4$};

    \draw (15.5,-3.0) -- (18.5,-3.0) -- (18.5,-4.0) -- (15.5,-4.0) -- cycle;
    \node at (16.2,-3.5) {$\overline \o_3$};

    \node at (17.8,-3.5) {${\o_3}$};

    \node[teal] at (15.7,2) {$F_2(\o_1)$};
    \node[blue] at (17.8,2) {$F_2(\overline{\o}_1)$};
    \node[orange] at (19.6,0.6) {$F_2(\o_2)$};
    \node[red] at (14.7,1.2) {$F_2(\o_4)$};

    \node at (3,-5) {(a)};
    \node at (10,-5) {(b)};
    \node at (17,-5) {(c)};

\end{tikzpicture}

\caption{ \footnotesize Two states connected by a colored line are in the same information set of $F_2$. In $(a)$, the $F_2$-sub-loops that
cover the $F_1$-loop are order-preserving, i.e., following the ordering of pairs in the original $F_1$-loop, whereas the sub-loop in $(c)$ is not order-preserving. 
(b) illustrates a case where $(\omega_1,\overline{\omega}_1, \overline{\omega}_3, \omega_3)$ forms an $F$-2 loop, 
but it is not an $F_2$-sub-loop of the original $F_1$-loop.} \label{Figure: covered by F2 loops}
\end{figure}

In simple terms, the definition states that, given an $F_1$-loop $(\omega_1,\overline{\omega}_1, \omega_2,\overline{\omega}_2, \dots, \omega_m,\overline{\omega}_m)$, we can partition its states into several $F_2$-loops and a set of states where $\o_j\in F_2(\overline{\omega}_j)$.
Figure \ref{Figure: covered by F2 loops} (a) depicts an $F_1$-loop consisting of $((\o_j,\overline{\o}_j))_{j=1,\dots,4}$, which is covered by two $F_2$-loops:  $(\o_1,\overline{\o}_1, \o_3,\overline{\o}_3)$ and
 $(\o_2,\overline{\o}_2, \o_4,\overline{\o}_4)$.
In this case, the set $J$ (defined in Definition \ref{Definition - covered loop}) is empty.
Figure \ref{Figure: covered by F2 loops} (b) depicts a case in which $J=\{2,4\}$, and $(\omega_1,\overline{\omega}_1, \overline{\omega}_3, \omega_3)$ forms an $F_2$-loop, 
yet it is not an $F_2$-sub-loop of the original $F_1$-loop since $\overline{\omega}_1$ is linked to $\overline{\omega}_3$ instead of $\omega_3$. 
Actually, if we set $A=\{\omega_2,\overline{\omega}_2,\omega_4,\overline{\omega}_4,\omega_1,\omega_3\}$ and 
$B=\{\overline{\omega}_1, \overline{\omega}_3\}$, which are $F_2$-measurable, then $\#(A\rightarrow B) =2$, but $\#(B\rightarrow A) =0$, 
so the $F_1$-loop is not $F_2$-balanced.   
Finally, note that the sub-loops in Figure \ref{Figure: covered by F2 loops} (a) are order-preserving. By contrast, the sub-loop $(\omega_1,\overline{\omega}_1,  \omega_3,\overline{\omega}_3, \omega_2,\overline{\omega}_2)$ in Figure \ref{Figure: covered by F2 loops} (c) 
does not preserve the ordering of the pairs as the $F_1$-loop, since the pair $(\omega_3,\overline{\omega}_3)$ appears before $(\omega_2,\overline{\omega}_2)$. 
In Section \ref{Section - necessary and sufficient conditions for dominance}, we show that order-preservation is needed to obtain a necessary condition for oracle dominance.

The following Proposition \ref{proposition: balanced} proves that an $F_1$-loop is $F_2$-balanced if and only if it is $F_2$-covered.
This proposition assists with the proof of Theorem \ref{Theorem: general case -- necessary condition} below, which provides a necessary condition for dominance.

\begin{proposition} \label{proposition: balanced}
   Let $(\omega_1,\overline{\omega}_1, \omega_2,\overline{\omega}_2 , \dots, \omega_m,\overline{\omega}_m)$ be an $F_1$-loop. The following statements are equivalent:
      \begin{description}
                 \item [i.] The loop is  $F_2$-balanced;
                 \item [ii.] The loop is $F_2$-covered;
                 \item [iii.] For every $F_2$-measurable   function  $f:\bigl\{\omega_1,\overline{\omega}_1, \omega_2,\overline{\omega}_2 , \dots, \omega_m,\overline{\omega}_m\bigr\} \to (0, \infty)$,
      $$
      \prod_{i=1}^m \frac{f(\omega_i)}{f(\overline{\omega}_i)}=1.
      $$  \end{description}
\end{proposition}

The next two properties that we study are \emph{irreducible} and \emph{informative} loops.
Starting with the former, an $F_i$-loop is irreducible if it does not have a \emph{sub-loop}, namely, there exists no `smaller' $F_i$-loop that comprises a strictly smaller set of states taken solely from the original loop.
Our analysis would use irreducible loops as building blocks to decompose and compare loops generated by the oracles' partitions.

\begin{definition}
    Let $L_i = (\omega_1, \overline{\omega}_1, \omega_2, \overline{\omega}_2, \dots, \omega_m, \overline{\omega}_m)$ be an $F_i$-loop.
    We say that the loop is \emph{irreducible} if there exists no strict subset of the set $\{\o_j,\overline{\o}_j: j=1,\dots,m\}$ that forms an $F_i$-loop.
\end{definition}

We use the definition of an irreducible loop in the context of covers as well, stating that a cover is \emph{irreducible} if every loop in the cover is irreducible.
Furthermore, the idea of irreducible loops is closely related to the concept of covers, and specifically to the set $J=\{j; \o_j\in F_{-i}(\overline{\omega}_j)\}$ given in Definition \ref{Definition - covered loop} above.
Specifically, if there exists an $F_i$-loop with a pair of states $(\o_j,\overline{\o}_j)$ such that $\overline{\o}_j \in F_i(\o_j)$, then it cannot be irreducible unless it comprises only $4$ states.\footnote{In general, the smallest possible loop has at least $4$ states, so any such loop is, by definition, irreducible.}
We typically refer to such cases where $\overline{\o}_j \in F_i(\o_j)$ as \emph{non-informative} because Oracle $i$ cannot distinguish between the two states.
This condition is essentially equivalent to every \( F_1 \)-loop being \( F_2 \)-balanced at $0$, meaning that for any choice of the specified \( F_2 \)-measurable sets \( A \) and \( B \), the number of transitions between these sets is zero.
The following Definition \ref{Definition: informative loops} captures the idea of \emph{informative} loops, which would later be used in Theorem \ref{Theorem: NI leads to dominance} as a sufficient condition for dominance.

\begin{definition} \label{Definition: informative loops}
     An $F_i$-loop $(\omega_1,\overline{\omega}_1, \omega_2,\overline{\omega}_2, \dots, \omega_m,\overline{\omega}_m)$ is $F_k$-\emph{non-informative} if $F_k(\omega_j) = F_k(\overline{\omega}_j)$ for every $j$.
     The loop is $F_k$-\emph{fully-informative} if $F_k(\omega_j) \neq  F_k(\overline{\omega}_j)$ for every $j$.
\end{definition}

To understand the motivation behind this definition, consider any $F_1$-loop denoted by $(\omega_1,\overline{\omega}_1, \omega_2,\overline{\omega}_2, \dots, \omega_m, \overline{\omega}_m)$.
If this loop is $F_2$-non-informative, it suggests that the ratios $\tfrac{\tau_2(s|\omega_i)}{\tau_2(s|\overline{\omega}_i)}$ equals $1$ for every signal $s$ supported on these states.
In simple terms, conditional on any $\{\omega_i,\overline{\omega}_i\}$, Oracle $2$ does not provide any additional information, so the constraints that an $F_1$-loop imposes on Oracle $1$ in every CKC (i.e., that the product of probability ratios along the loop equals $1$) are met by the measurability requirements of $F_2$.

The following proposition summarizes key properties of informative and irreducible loops.
It states that an irreducible loop intersects every CKC at most once and must be fully informative (unless it has only 4 states).
In addition, the proposition shows that an informative loop has a fully-informative sub-loop, as well.

\begin{proposition} \label{Proposition - irreducible and informative loops}
Consider an $F_i$-loop $L_i$.
\begin{itemize}
    \item If $L_i$ intersects the same \emph{CKC} more than once, then it is not irreducible.
    \item If $L_i$ is irreducible and consists of at least $6$ states, then it is $F_i$-fully-informative.
    \item If $L_i$ is $F_i$-informative, then it has an $F_i$-fully-informative sub-loop.
    \item If $L_i$ is $F_i$-fully-informative, then it can be decomposed to irreducible $F_i$-loops.
    \item If $L_i$ is not irreducible, then either it intersects the same \emph{CKC} more than once, or it has at least 4 states in the same partition element of $F_i$.
\end{itemize}
\end{proposition}

We use this proposition in the following subsection to provide necessary and sufficient conditions for the dominance of one oracle over another.

\section{Necessary and Sufficient conditions for dominance} \label{Section - necessary and sufficient conditions for dominance}

In the following section, we address the general case where \( F_1 \) has loops, which imposes constraints on Oracle $1$ \emph{across} CKCs.
Due to the complexity of this problem, we divide our analysis into two parts: a necessary condition for dominance presented in Theorem \ref{Theorem: general case -- necessary condition}, and a sufficient condition given in Theorem \ref{Theorem: NI leads to dominance}.
These theorems depend strongly on the properties of information loops, and specifically on the notions of covers, irreducibility and non-informativeness.

Starting with the necessary conditions, the following theorem, which builds on Propositions \ref{proposition: balanced} and \ref{Proposition - irreducible and informative loops}, states that if Oracle $1$ dominates Oracle $2$, then besides the refinement condition in every CKC, already established in Theorem \ref{Theorem - no loop characterization}, it must be that every $F_1$-loop is covered by loops of $F_2$.
In addition, it states that every irreducible $F_2$-loop that covers an irreducible $F_1$-loop is order-preserving, essentially stating that the two loops coincide.

\begin{theorem} \label{Theorem: general case -- necessary condition}
If \emph{Oracle} $1$ dominates \emph{Oracle} $2$, then:
\begin{itemize}
    \item $F_1$ refines $F_2$ in every \emph{CKC};
    \item Any $F_1$-loop has a cover by $F_2$-loops; and
    \item Every irreducible $F_2$-loop that covers an irreducible $F_1$-loop is order-preserving.
\end{itemize}
\end{theorem}

The proof of the first part is immediate, as it follows directly from the main result of Part I (see Theorem $4$ therein cited in Appendix \ref{Subsection - a unique CKC}).
The proof of the second part relies on Proposition \ref{proposition: balanced} by assuming that an $F_1$-loop is not $F_2$-balanced, and constructing a strategy $\tau_2$ that Oracle $1$ cannot mimic without violating measurability constraints.
The last part relies on Proposition \ref{Proposition - irreducible and informative loops}, as well as a key lemma from Part I (cited in Appendix \ref{Subsection - proportional signals with m states}), by depicting a two-signal strategy $\tau_2$ that one cannot mimic without following the same order of pairs throughout the $F_2$-loop.

Next, we use the understanding regarding covered and balanced loops to present a sufficient condition for dominance, which indirectly requires that any loop is balanced at $0$—meaning that there are no transitions between sets \( A \) and \( B \).
This leads to the following Theorem \ref{Theorem: NI leads to dominance}, which uses the non-informative notion for dominance.

\begin{theorem} \label{Theorem: NI leads to dominance}
If $F_1$ refines $F_2$ in every \emph{CKC} and every $F_1$-loop is $F_2$-non-informative, then \emph{Oracle} $1$ dominates \emph{Oracle} $2$.
\end{theorem}

Though we do not yet provide a full characterization, it becomes rather clear that the requirement that every $F_1$-loop is $F_2$-balanced should be the main focus, as it is a necessary condition, as well as a sufficient one when the balance is set to zero.
In the following section we show that the balance condition is both necessary and sufficient for the case of two CKCs.

\subsection{The case of two CKCs} \label{Section - Toward a general characterization: two CKCs}

In this section, we assume there are only two CKCs.
This assumption simplifies the analysis, as the case of two CKCs can be resolved using our prior results, allowing us to examine all possible loops directly.
Formally, Proposition \ref{Proposition: two CKCs} states that, given two CKCs, the necessary condition of an \( F_2 \)-balanced loop from Theorem \ref{Theorem: general case -- necessary condition} is also a sufficient condition.

To build intuition, consider the scenario with two CKCs depicted in Figure \ref{fig: two CKCs and balanced loops}, featuring an \( F_1 \)-loop \((\omega_1, \overline{\omega}_1, \omega_2, \overline{\omega}_2)\) across four states.
Fix some \(\tau_2\) and assume the loop is \( F_2 \)-balanced.
There are then only two possibilities: either the loop is \( F_2 \)-non-informative, as shown in cases (a) and (b) in Figure \ref{fig: two CKCs and balanced loops}, or it is also an \( F_2 \)-loop, illustrated in case (c) in Figure \ref{fig: two CKCs and balanced loops}.
The first possibility was covered in Theorem \ref{Theorem: NI leads to dominance}, while the second allows Oracle $1$ to meet the constraints imposed by the \( F_1 \)-loop when attempting to mimic \(\tau_2\).

\begin{proposition} \label{Proposition: two CKCs}
    Assume there are only two \emph{CKCs}. Then, \emph{Oracle} $1$ dominates \emph{Oracle} $2$ if and only if $F_1$ refines $F_2$ in every \emph{CKC} and any $F_1$-loop is $F_2$-balanced.
\end{proposition}

\begin{figure}[th!]
\centering
\begin{minipage}{0.3\textwidth}
\centering
\begin{tikzpicture}[scale=0.7]

\draw[thick] (0,0) rectangle (6,6);

\draw[black, thick] (3,0) -- (3,6);

\node at (1.6,5.6) {$C_1$};
\node at (4.6,5.6) {$C_2$};

\node[blue] at (1.2,2.8) {$F_2$};

\draw[blue, thick] (1.6,2.8) ellipse (1 and 2.3);
\draw[blue, thick] (4.6,2.8) ellipse (1 and 2.3);

\filldraw[black] (1.4,4.3) circle (2pt) node[anchor=west] {$\omega_1$};
\filldraw[black] (1.4,1.3) circle (2pt) node[anchor=west] {$\overline{\omega}_1$};
\filldraw[black] (4.4,4.3) circle (2pt) node[anchor=west] {$\overline{\omega}_2$};
\filldraw[black] (4.4,1.3) circle (2pt) node[anchor=west] {$\omega_2$};

\end{tikzpicture}
\caption*{(a)}
\vspace{-0.2cm}
\end{minipage}%
\hspace{0.3cm}
\begin{minipage}{0.3\textwidth}
\centering
\begin{tikzpicture}[scale=0.7]

\draw[thick] (0,0) rectangle (6,6);

\draw[black, thick] (3,0) -- (3,6);

\node at (1.6,5.6) {$C_1$};
\node at (4.6,5.6) {$C_2$};

\node[blue] at (1.2,2.8) {$F_2$};

\draw[blue, thick, rounded corners=0.5cm] (0.7,0.5) rectangle (5.6,5);

\filldraw[black] (1.4,4.3) circle (2pt) node[anchor=west] {$\omega_1$};
\filldraw[black] (1.4,1.3) circle (2pt) node[anchor=west] {$\overline{\omega}_1$};
\filldraw[black] (4.4,4.3) circle (2pt) node[anchor=west] {$\overline{\omega}_2$};
\filldraw[black] (4.4,1.3) circle (2pt) node[anchor=west] {$\omega_2$};

\end{tikzpicture}
\caption*{(b)}
\vspace{-0.2cm}
\end{minipage}%
\hspace{0.3cm}
\begin{minipage}{0.3\textwidth}
\centering
\begin{tikzpicture}[scale=0.7]

\draw[thick] (0,0) rectangle (6,6);

\draw[black, thick] (3,0) -- (3,6);

\node at (1.6,5.6) {$C_1$};
\node at (4.6,5.6) {$C_2$};

\node[blue] at (1.2,2.8) {$F_2$};

\draw[blue, thick] (3,1.2) ellipse (2.5 and 0.8);
\draw[blue, thick] (3,4.2) ellipse (2.5 and 0.8);

\filldraw[black] (1.4,4.3) circle (2pt) node[anchor=west] {$\omega_1$};
\filldraw[black] (1.4,1.3) circle (2pt) node[anchor=west] {$\overline{\omega}_1$};
\filldraw[black] (4.4,4.3) circle (2pt) node[anchor=west] {$\overline{\omega}_2$};
\filldraw[black] (4.4,1.3) circle (2pt) node[anchor=west] {$\omega_2$};

\end{tikzpicture}
\caption*{(c)}
\vspace{-0.2cm}
\end{minipage}%
\caption{Two CKCs with an $F_1$-loop described by $(\omega_1,\overline{\omega}_1,\omega_2,\overline{\omega}_2)$. Graph $(a)$ and $(b)$ depict two $F_2$-balanced loops, that are also $F_2$-non-informative, and $(c)$ describes an $F_2$-loop. Any other structure of $F_2$ yields a non-balanced loop.}
\label{fig: two CKCs and balanced loops}
\end{figure}

\section{Equivalent oracles} \label{Section - Equivalent oracles}

In this section we tackle a parallel question to dominance, which is the problem of oracles' equivalence.
Specifically, we characterize necessary and sufficient conditions such that both oracles dominate one another simultaneously, as formally given in the following definition:

\begin{definition} \label{Definition - equivalent oracles}
$F_1$ is \emph{equivalent} to $F_2$, denoted $F_1\sim F_2$, if the two oracles dominate one another, that is, if $F_i\succeq_{\rm{NE}} F_{-i}$ for every $i=1,2$.
\end{definition}

Based on the results for the case that loops do not exist and the case of two CKCs, equivalence between oracles obviously requires two-sided refinement within every CKC (i.e., equivalence), and that every $F_i$-loop is $F_{-i}$-balanced for every Oracle $i$.
This, however, is insufficient and equivalence also requires that every irreducible $F_i$-loop with at least $6$ states is also an irreducible $F_{-i}$-loop.
This result is given in the following Theorem \ref{Theorem: Equivalent oracles}.

\begin{theorem}\label{Theorem: Equivalent oracles}
$F_1$ is \emph{equivalent} to $F_2$ if and only if for every \emph{Oracle} $i$, the partition $F_i$ refines $F_{-i}$ in every \emph{CKC}, any $F_i$-loop has a cover of $F_{-i}$-loops, and every irreducible $F_i$-loop with at least $6$ states is an irreducible $F_{-i}$-loop.
\end{theorem}

The equivalence condition concerning irreducible loops is based on the ability of both oracles to follow similar measurability constraints when signaling to players in every CKC.
That is, if one oracle is constrained by an information loop, then we require the other to follow suit.
Yet, this still raises the question of why we need to focus on irreducible loops.
To understand this, consider a single partition element of $F_i$ that intersects at least two CKCs where each intersection contains at least two states.
This evidently generates a non-informative loop, because all pairs are non-informative.
But as long as the other oracle cannot distinguish between the two states in each pair, the ability to separate different pairs in different CKCs is not needed, as each pair is common knowledge among the players themselves within every CKC.

The proof of Theorem \ref{Theorem: Equivalent oracles} also builds on an intermediate irreducibility notion that we refer to as \emph{type-2 irreducible loop}.
More formally, an $F_i$-loop is type-2 irreducible if it does not have four states from the same partition element of $F_i$.
This notion refines that of fully-informative loops (as every type-2 irreducible loop is fully-informative), but also weakens that of irreducible loops, because a type-2 irreducible loop can intersect the same CKC multiple times, and so be decomposed into sub-loops.

The notion of type-2 irreducible loops is crucial for our analysis and results, but also in a more general manner.
We use type-2 irreducible loops to generate the basic elements, \emph{building blocks}, upon which two oracles must match one another (in terms of their information).
These building blocks are referred to as \emph{clusters} and they are constructed as follows.
First, we take the set of type-2 irreducible loops.
Then, we consider such loops that intersect the same CKC and consider them as connected.
Next, we take the transitive-closure of this relation, which yield disjoint sets of connected type-2 irreducible loops.
Finally, we take every such set (of connected loops) and consider all the CKCs that it intersects; this is a cluster.
We prove that the oracles' partitions match one another in each of these clusters.
That is, the clusters are the basic structure upon which we derive an equivalence, and later extend it to ``simpler" connections between clusters that involve only a single partition element of $F_i$.

\newpage


\bibliographystyle{chicago}

\bibliography{refs.bib}


\appendix

\section{Appendices}

\subsection{Key results from the companion Part I}\label{Appendix_Part I results}

\subsubsection{Proportionality lemma from Part I} \label{Subsection - proportional signals with m states}

Fix two distinct signals $\{s_1,s_2\}$ and assume that the partition $F_2=\{A_1,A_2,\dots,A_m\}$ has $m$ elements, as noted.
Let $p_1,p_2,\dots,p_m$ be $m$ distinct probabilities such that all ratios of two distinct numbers from the set $\mathbb{A} = \{p_j,1-p_j:j=1,2,\dots,m\}$ are pairwise different.\footnote{To achieve this, one can consider $m$ distinct prime numbers $r_1<r_2<\dots<r_m$. Define $\mathbb{T}_0=\mathbb{Q}$, and for every $j\geq 1$, let $\mathbb{T}_j$ be the extended field of $\mathbb{T}_{j-1}$ with $\sqrt{r_{j}}$. Take $p_j \in \mathbb{T}_j \setminus \mathbb{T}_{j-1}$.}
Denote by $ {\rm Post}(\tau_i)$ the set of posteriors induced by the strategy $\tau_i$. 
Define the signaling function $\tau_2$ such that
\begin{equation} \label{Equation - fully support tau_2 on m elements}
    \tau_2(s_1|A_j)=1-\tau_2(s_2|A_j)=p_j, \ \ \forall \leq j \leq m.
\end{equation}
Given this signaling function and assuming that the state space comprises a unique CKC, Lemma \ref{Lemma - proportional signals with m states} (from Part I) states that the condition ${\rm Post}(\tau_1) \subseteq {\rm Post}(\tau_2)$ implies that $\tau_1$ is partially proportional to $\tau_2$, restricted to a subset of feasible signals.

\begin{lemma} \label{Lemma - proportional signals with m states}
    Fix $\tau_2$ given in \emph{Equation} \eqref{Equation - fully support tau_2 on m elements} and a unique \emph{CKC}.
If ${\rm Post}(\tau_1) \subseteq {\rm Post}(\tau_2)$, then for every signal $t \in {\rm Supp}(\tau_1)$ there exists a signal $s \in \{s_1,s_2\}$ and a constant $c>0$ such that $\tau_1(t|\omega)=c \tau_2(s|\omega)$ for every $\omega \in\Omega$.
\end{lemma}

\subsubsection{Unique CKC, characterization result from Part I}  \label{Subsection - a unique CKC}

\begin{theorem} \label{Theorem - a unique CKC}
    Assume that $\Omega$ comprises a unique common knowledge component.
    Then, the following are equivalent:
    \begin{itemize}
        \item $F_1$ refines $F_2$;
        \item $ F_1 \succeq_{\rm{NE}} F_2$;
        \item For every $\tau_2$, there exists $\tau_1$, so that ${\rm Post}(\tau_1) \subseteq {\rm Post}(\tau_2)$;
        \item For every $\tau_2$, there exists $\tau_1$, so that ${\rm Post}(\tau_1) = {\rm Post}(\tau_2)$;
        \item For every $\tau_2$, there exists $\tau_1$, so that $\mu_{\tau_1} = \mu_{\tau_2}$.
    \end{itemize}
\end{theorem}

\subsection{Proof of Theorem \ref{Theorem - no loop characterization}}

\begin{proof}
One direction is straightforward.
 Assume, to the contrary, that  Oracle $1$ dominates Oracle $2$, but $F_1$ does not refine $F_2$ in some CKC.
Denote this CKC by $C_1$, and consider the set of all games in which the payoffs of all players are zero in every $\o \notin C_1$, independent of their actions.
Thus, Oracle $1$ dominates Oracle $2$ in every game restricted to $C_1$, although $F_1$ does not refine $F_2$ in $C_1$.
This contradicts the ket result from Part~I (see Theorem \ref{Theorem - a unique CKC} in Appendix \ref{Subsection - a unique CKC}).

Moving on to the second part, assume to the contrary that $F_1$ refines $F_2$ in every CKC, but Oracle $1$ does not dominate Oracle $2$.
Therefore, there exists a strategy $\tau_2$ such that Oracle $1$ cannot produce the same distribution over posteriors as $\tau_2$.\footnote{Observe that
the condition that Oracle 1 can generate the same distribution over posterior profiles as Oracle 2 implies that Oracle 1 dominates Oracle 2.
To see this, consider any game and any signaling strategy $\tau$. Since the players' strategies depend on the profile of posteriors,
we can then abstract away from the underlying private and public information and assume that the players play a Bayesian both Oracles can generate distributions over the profiles of posteriors, which can be generated by both Oracles. }
The proof now splits to $4$ steps.

\noindent \textbf{Step 1: Mimicking sub-strategies.}

We start by defining the notion of a sub-strategy, which resembles a strategy, but with induced probabilities that may sum to less than $1$.
Formally, a \emph{partial distribution} $\tilde{p}$ is a non-negative function from a finite subset of $S$ to $[0,1]$ such that $\sum_{s\in S}\tilde{p}(s) \leq 1$.
A partial distribution differs from a distribution as the probabilities need not sum to $1$.
Let $\tilde{\Delta}(S)$ be the set of partial distributions on $S$, and define a \emph{sub-strategy} $\underline{\tau} : \Omega \to \tilde{ \Delta }(S)$ as an $F_1$-measurable function from $\Omega$ to the set of partial distributions on $S$.
That is, $\underline{\tau}(s|\o)\geq 0$ and $\sum_{s}\underline{\tau}(s|\o) \leq 1$, for every $\o$ and $s$.
Evidently, every $F_1$-measurable strategy is a sub-strategy.

For every sub-strategy $\underline{\tau}$ and every $p \in (\Delta(\Omega))^n$, let $\mathbf{P}_{\underline{\tau}}(p)$ be the probability that $\underline{\tau}$ yields the posterior $p$, i.e.,
\begin{equation} \label{Posterior measure}
\mathbf{P}_{\underline{\tau}}(p) = \sum_{\substack{(\o,s): \ \underline{\tau}(s|\o)>0, \\ {\rm \ and \ }(\mu^i_{\underline{\tau}|\o,s})_{i\in N}=p}}\mu(\o)\underline{\tau}(s|\o).
\end{equation}
Similarly, define $\mathbf{P}_{\tau_2}(p)$ for every posterior $p$ given the stated strategy $\tau_2$.
We say that a sub-strategy $\underline{\tau}$ \emph{mimics} $\tau_2$ if
\begin{equation} \label{Inequality of posteriors mimicking}
    \mathbf{P}_{\underline{\tau}}(p) \leq \mathbf{P}_{\tau_2}(p), \text{ for every } p\in (\Delta(\Omega))^n.
\end{equation}
Hence, a sub-strategy $\underline{\tau}$ mimics $\tau_2$ if, for every posterior $p$, the probability that $\underline{\tau}$ generates $p$ does not exceed the probability that $\tau_2$ generates it.
Note that the null sub-strategy (i.e., $\underline{\tau}(s|\o)=0$ for every $\o$ and $s$) also mimics $\tau_2$.

Consider any sub-strategy $\underline{\tau}$ that mimics $\tau_2$.
Because $\tau_2$ generates a finite set ${\rm Post}(\tau_2)$ of possible posteriors, there exists a finite number of combinations of posteriors (which does not exceed $2^{|{\rm Post}(\tau_2)|}$) that every signal of $\underline{\tau}$ supports.
So, if some sub-strategy uses more than $2^{|{\rm Post}(\tau_2)|}$ signals, we can apply the pigeonhole principle to deduce that the additional signals support similar combinations of posteriors as some other signals.
Therefore, for every such additional signal $s$, there exists another signal $s'$ and a constant $c>0$ such that $\underline{\tau}(s|\o) = c  \underline{\tau}(s'|\o)$ for every $\o$, and we can unify the two signals into one.
We can thus assume that there exists a finite set of signals $\underline{S}$, such that every mimicking sub-strategy (i.e., that mimics $\tau_2$) uses only signals from $\underline{S}$.

\bigskip

\noindent \textbf{Step 2: Optimal sub-strategies.}

Let $A_{\underline{\tau}}$ be the set of sub-strategies that mimic $\tau_2$.
Note that the set of sub-strategies supported on $\underline{S}$ is compact, and the (inequality) mimicking condition, $\mathbf{P}_{\underline{\tau}}(p) \leq \mathbf{P}_{\tau_2}(p)$ for every $p\in (\Delta(\Omega))^n$, remains valid when considering a converging sequence of sub-strategies.
Thus, $A_{\underline{\tau}}$ is also compact.

Consider the function $H(\underline{\tau}) = \sum_{p\in {\rm Post}(\tau_2)}\mathbf{P}_{\underline{\tau}}(p)$ defined from $A_{\underline{\tau}}$ to $[0,1]$.
As a piece-wise linear function of $\tau$, it is a continuous, so $\underline{\tau}_{1.0} = {\argmax_{\underline{\tau} \in A_{\underline{\tau}}}} H(\underline{\tau})$ is well-defined.
If $H(\underline{\tau}_{1.0})=1$, then $\underline{\tau}_{1.0}$ is an $F_1$-measurable strategy that mimics $\tau_2$.
This contradicts the original premise (that Oracle $1$ cannot induce the same distribution over posteriors as $\tau_2$), so assume to the contrary that $\underline{\tau}_{1.0}$ is a proper sub-strategy and $H(\underline{\tau}_{1.0})<1$.
If that is the case (i.e., if $H(\underline{\tau}_{1.0})<1$), there exists a posterior $p^* \in {\rm Post}(\tau_2)$ so that $\mathbf{P}_{\underline{\tau}_{1.0}}(p^*) < \mathbf{P}_{\tau_2}(p^*)$.

\bigskip

\noindent \textbf{Step 3: Partially supported and connected posteriors.}

For every posterior $p \in {\rm Post}(\tau_2)$, let $A_{p} =\{\o \in \Omega: \ p^{i}(\o)>0  \text{ for some player } i \} $ be the set of states on which $p$ is strictly positive, contained in some CKC denoted $C_{p}$.
We say that a posterior $p \in {\rm Post}(\tau_2)$ is \emph{partially supported} (PS) if
$\mathbf{P}_{\underline{\tau}_{1.0}}(p) < \mathbf{P}_{\tau_2}(p)$, otherwise we say that $p$ is \emph{fully supported} (FS).
Let us now prove a few supporting claims related to PS posteriors.


\bigskip

\noindent \textbf{Claim 1:} If $p$ is PS, then $\sum_{s}\underline{\tau}_{1.0}(s|\o)<1$ for every state $\o\in A_p$.

\begin{proof}
Fix a posterior $p$ and a state $\o_0$ such that $(\mu^i_{\tau|\o_0,s})_{i\in N}= p$ for some signal $s$ and $\tau \in \{\underline{\tau}_{1.0}, \tau_2 \}$.
There exists a constant $\alpha_{p,\o_0}$, independent of $s$ and $\tau$, such that $\alpha_{p,\o_0} \mu(\o_0) \tau(s|\o_0) = \sum_{\o \in A_p\setminus\{\o_0 \}} \mu(\o) \tau(s|\o) $.
This follows from the fact that, in order to induce the posterior $p$, the probabilities induced by $\tau$ must maintain the same proportions along the different states in $A_p$, independently of either the strategy or the signal.
Otherwise, the induced posterior would not match $p$.
Thus, Equation \eqref{Posterior measure} could be re-formulated as follows,
\begin{eqnarray*}
    \mathbf{P}_{{\tau}}(p)
        & = & \sum_{(\o,s):(\mu^i_{{\tau}|\o,s})_{i\in N}=p}\mu(\o){\tau}(s|\o) \\
        & = &  \sum_{s:(\mu^i_{{\tau}|\o_0,s})_{i\in N}=p}  \mu(\o_0) {\tau}(s|\o_0) +
        \sum_{\substack{(\o,s): \o \in A_p\setminus\{\o_0\}, \\ {\rm \ and \ }(\mu^i_{{\tau}|\o,s})_{i\in N}=p}}\mu(\o){\tau}(s|\o) \\
        & = & (1+\alpha_{p,\o_0}) \mu(\o_0) \sum_{s:(\mu^i_{{\tau}|\o_0,s})_{i\in N}=p}   {\tau}(s|\o_0),
\end{eqnarray*}
which translates to
$$
\sum_{s:(\mu^i_{{\tau}|\o_0,s})_{i\in N}=p}   {\tau}(s|\o_0) = \frac{\mathbf{P}_{{\tau}}(p) }{(1+\alpha_{p,\o_0}) \mu(\o_0)}.
$$
Summing over all $p \in {\rm Supp}(\tau_2)$, we get
\begin{equation} \label{Equation - posterior measure sum s}
    \sum_{s}{\tau}(s|\o_0) = \frac{1}{\mu(\o_0)}\sum_{\substack{p : (\mu^i_{{\tau}|\o_0,s})_{i\in N}=p, \\ {\rm \ for \ some \ } s}}\frac{\mathbf{P}_{{\tau}}(p) }{(1+\alpha_{p,\o_0})}.
\end{equation}
Note that the RHS holds for either $\underline{\tau}_{1.0}$ or $\tau_2$.

Now assume, by contradiction, that $p_0$ is a PS posterior and $\sum_{s}\underline{\tau}_{1.0}(s|\o_0)=1$ for some state $\o_0\in A_{p_0}$.
Using Equation \eqref{Equation - posterior measure sum s}, for both $\tau_2$ and $\underline{\tau}_{1.0}$, we get
\begin{eqnarray*}
    1 & = &  \sum_{s}{\tau_2}(s|\o_0) = \frac{1}{\mu(\o_0)}\sum_{\substack{p : (\mu^i_{{\tau_2}|\o_0,s})_{i\in N}=p, \\ {\rm \ for \ some \ } s}}\frac{\mathbf{P}_{{\tau_2}}(p) }{(1+\alpha_{p,\o_0})}  \\
    1  & = &  \sum_{s}\underline{\tau}_{1.0}(s|\o_0)  = \frac{1}{\mu(\o_0)}\sum_{\substack{p : (\mu^i_{\underline{\tau}_{1.0}|\o_0,s})_{i\in N}=p, \\ {\rm \ for \ some \ } s}}\frac{\mathbf{P}_{\underline{\tau}_{1.0}}(p) }{(1+\alpha_{p,\o_0})},
\end{eqnarray*}
which implies that
$$
\sum_{\substack{p : (\mu^i_{{\tau_2}|\o_0,s})_{i\in N}=p, \\ {\rm \ for \ some \ } s}}\frac{\mathbf{P}_{{\tau_2}}(p) }{(1+\alpha_{p,\o_0})} = \sum_{\substack{p : (\mu^i_{\underline{\tau}_{1.0}|\o_0,s})_{i\in N}=p, \\ {\rm \ for \ some \ } s}}\frac{\mathbf{P}_{\underline{\tau}_{1.0}}(p) }{(1+\alpha_{p,\o_0})}  < \sum_{\substack{p : (\mu^i_{{\tau_2}|\o_0,s})_{i\in N}=p, \\ {\rm \ for \ some \ } s}}\frac{\mathbf{P}_{{\tau_2}}(p) }{(1+\alpha_{p,\o_0})},
$$
where the strict inequality follows from the fact that $\mathbf{P}_{\underline{\tau}_{1.0}}(p)  \leq \mathbf{P}_{\tau_2}(p)$ for every posterior $p$, with a strict inequality for $p=p_0$.
This yields a contradiction, and the result follows.
    \hfill
\end{proof}

\noindent \textbf{Claim 2:} If $\sum_{s}\underline{\tau}_{1.0}(s|\o)<1$ for some state $\o$, then there exists a PS posterior $p$ such that $\o \in A_p$.

\begin{proof}
Assume, to the contrary, that $\sum_{s}\underline{\tau}_{1.0}(s|\o_0)<1$ for some state $\o_0$, and every posterior $p$ such that $\o_0 \in A_p$ is FS.
Using Equation \eqref{Equation - posterior measure sum s}, we deduce that
\begin{eqnarray*}
    1
    & = &  \sum_{s}{\tau_2}(s|\o_0)  \\
    & = & \frac{1}{\mu(\o_0)}\sum_{\substack{p : (\mu^i_{{\tau_2}|\o_0,s})_{i\in N}=p, \\ {\rm \ for \ some \ } s}}\frac{\mathbf{P}_{{\tau_2}}(p) }{(1+\alpha_{p,\o_0})}  \\
    & = &  \frac{1}{\mu(\o_0)}\sum_{\substack{p : (\mu^i_{\underline{\tau}_{1.0}|\o_0,s})_{i\in N}=p, \\ {\rm \ for \ some \ } s}}\frac{\mathbf{P}_{\underline{\tau}_{1.0}}(p) }{(1+\alpha_{p,\o_0})} \\
    & = & \sum_{s}\underline{\tau}_{1.0}(s|\o_0) < 1,
\end{eqnarray*}
where the first equality follows from the fact that $\tau_2$ is a strategy, the second and fourth equations follow from Equation \eqref{Equation - posterior measure sum s}, the third equality follows from the fact that every posterior $p$ such that $\o_0 \in A_p$ is FP, and the last inequality is by assumption.
We thus reach a contradiction, and the result follows.
    \hfill
\end{proof}

We will use Claims $1$ and $2$ to extend $\underline{\tau}_{1.0}$, and show that it cannot be a maximum of $H$.
For this purpose we need to define the notion of connected posteriors.
Formally, we say that two posteriors $p,p' \in {\rm Post}(\tau_2)$ are \emph{connected} if there exist two states $(\o,\o') \in A_p\times A_{p'} \subseteq C_{p} \times C_{p'} $, where  $C_{p} \neq C_{p'}$ are two distinct CKCs, such that $F_1(\o)=F_1(\o')$.
Equivalently, in such a case, we refer to $C_p$ and $C_{p'}$ as \emph{connected}, as well.
Let $(\o,\o')$ and $F_1(\o)$ be the \emph{connection} and \emph{connecting set} of $p$ and $p'$, respectively.\footnote{Equivalently, we refer to $(\o,\o')$ and $F_1(\o)$ as the \emph{connection} and \emph{connecting set} of the CKCs $C_{p}$ and $C_{p'}$.}
We can now relate the notion of connected posteriors to PS ones through the following claim.

\bigskip

\noindent \textbf{Claim 3:} Fix a PS posterior $p$ and $\o \in A_p$. Then, for every connection $(\o,\o')$, there exists a PS posterior $p'$ such that $\o' \in A_{p'} \cap F_1(\o)$.

\begin{proof}
Let $p$ be a PS posterior with a connection $(\o,\o')$ and $F_1(\o) =F_1(\o')$.
Using Claim $1$, if $p$ is PS, then $\sum_{s}\underline{\tau}_{1.0}(s|\o)<1$ for every $\o \in A_p$, so the $F_1$-measurability constraint implies that $\sum_{s}\underline{\tau}_{1.0}(s|\o')<1$.
Thus, according to Claim $2$, there exists a PS posterior $p'$ such that $\o' \in A_{p'}$, as needed.
    \hfill
\end{proof}

\noindent \textbf{Step 4: Extending $ \underline{\tau}_{1.0}$.}

Recall that $p^*$ is a PS posterior.
Let $V$ be the set of all CKCs $C_{l}$ such that there exists a sequence of PS posteriors $(p^*,p_1,\dots,p_l)$ where every two successive posteriors are connected and $A_{p_l} \subseteq C_{l}$.
Assume that $V$ also contains $C_{p^*}$.
Let $E \subseteq V^2$ be the set of couples $(C,C')$ such that $C$ and $C'$ are connected, and denote by $\mathcal{P}^*$ the set of all PS connected posteriors that generate $V$.
Clearly, $(V,E)$ is a connected graph and we can use it to construct a sub-strategy $\underline{\tau}$ which mimics $\tau_2$ and  ${\rm Post}(\underline{\tau}) = \mathcal{P}^*$.
The proof proceeds by induction on the number of vertices in $V$.

\bigskip

\noindent \textbf{Preliminary step: $|V|=1$.}
Assume that $C_{p^*}$ is the unique CKC in $V$.
Because $p^* \in {\rm Post}(\tau_2)$, there exists a signal $s^*$ and state $\o \in C_{p^*}$ such that $\tau_2(s^*|\o)>0$ and $(\mu^i_{\tau_2|\o,s^*})_{i\in N}=p^*$.
Define the sub-strategy
$\underline{\tau}_{1.1}(s|\omega) = \tau_2(s^*|\omega)$ for every $\o \in A_{p^*}$.
Recall that $F_1$ refines $F_2$ in every CKC, therefore $\underline{\tau}_{1.1}$ is well defined.
Moreover, it is a sub-strategy that mimics $\tau_2$ and ${\rm Post}(\underline{\tau}_{1.1}) = \mathcal{P}^*$, as needed.

\bigskip

\noindent \textbf{Induction step: $|V|=m$.}
Assume that for every graph $(V,E)$ where $|V| = m$, there exists a sub-strategy $\underline{\tau}_{1.m}$ that mimics $\tau_2$, and ${\rm Post}(\underline{\tau}_{1.m}) = \mathcal{P}^*$.

\bigskip

\noindent \textbf{Induction proof for $|V|=m+1$.}
Assume that $|V|=m+1$.
The distance between $C_{p^*}$ and every vertex (i.e., every CKC) in $V$ is defined by the shortest path between the two vertices.
Denote by $C_{m+1}$ the vertex in $(V,E)$ with the longest path from $C_{p^*}$.

We argue that $C_{m+1}$  has exactly one connecting set with the other vertices.
Otherwise, assume that there are at least two connecting sets.
If the two originate from the same CKC in $V$, then we get an $F_1$-loop, which cannot exist.
Thus, we can assume that the two sets originate from different CKCs, denoted $C$ and $C'$.
Since $(V,E)$ is a connected graph, there exists a path from $C_{p^*}$ to each of these CKCs.
Consider the two sequences of connecting sets for these two paths.
If the two are pairwise disjoint, then we have an $F_1$-loop from $C_{p^*}$ to $C_{m+1}$, which again yields a contradiction.
So the sequences must coincide at some stage.
Take a truncation of the sequences from the last stage in which they coincide until $C_{m+1}$.
The origin of the two paths are connected CKCs (sharing the same connecting set), denoted $C_l$ and $C_{l+1}$, so we now have two pairwise disjoint sequences between these two connected CKCs till $C_{m+1}$, thus generating an $F_1$-loop.
Therefore, we conclude that there is exactly one connecting set, denoted $A$, between $C_{m+1}$ and the other CKCs in $V$.

Consider a refinement of $F_1$ where $A$ is partitioned into two disjoint sets, $A_1 = A \setminus C_{m+1}$ and $A_2 = A \cap C_{m+1}$.
In such a case, $|V|=m$ and, according to the induction step, there exists a mimicking sub-strategy $\underline{\tau}_{1.m}$ supported on every PS connected posterior in $\mathcal{P}^*$ other than the ones related to the CKC $C_{m+1}$.
Let $p_{m+1}$ denote a PS posterior such that $A_2 \subset A_{p_{m+1}} \subseteq C_{m+1}$.
In case there is more than one PS posterior, the proof works similarly because every additional posterior shares the same connecting set $A$.

According to the induction step, ${\rm Post}(\underline{\tau}_{1.m}) = \mathcal{P}^*\setminus \{p_{m+1}\}$, so we need to extend this sub-strategy to support $p_{m+1}$ as well.
Since $p_{m+1} \in {\rm Post}(\tau_2)$, there exists a signal, denoted $s^*$ w.l.o.g., and states $\o \in A_{p_{m+1}} \subseteq C_{m+1}$ such that $\tau_2(s^*|\o)>0$ and $(\mu^i_{\tau_2|\o,s^*})_{i\in N}=p_{m+1}$.
Moreover, because $C_{m+1}$ is not connected (neither directly, nor indirectly)  to the other CKCs in $V$ under the refined $F_1$, we can assume that $\sum_s \underline{\tau}_{1.m}(s|A_1) > \sum_s \underline{\tau}_{1.m}(s|A_2)$. Otherwise, we can re-scale $\underline{\tau}_{1.m}$ in the different \emph{unconnected} elements of the refined $F_1$.
Hence, we can also assume that there exists a signal, again denoted $s^*$ w.l.o.g., such that $\underline{\tau}_{1.m}(s^*|A_1) > 0 =  \underline{\tau}_{1.m}(s^*|A_2)$.

Define the following function
\begin{equation*}
    \underline{\tau}_{1.m+1}(s|\omega) =
    \begin{cases}
        c_m \underline{\tau}_{1.m}(s|\omega), & \text{for every } (\omega, s) \ \text{s.t. } \underline{\tau}_{1.m}(s|\o)>0, \\
        c_{2} \underline{\tau}_{2}(s^*|\omega), & \text{for every }  (\omega, s) \ \text{s.t. } \o \in A_{p_{m+1}}, \ s=s^*,
    \end{cases}
\end{equation*}
where the parameters $c_m>0$ and $c_2>0$ are chosen to ensure that  $\underline{\tau}_{1.m+1}(s^*|A_1) = \underline{\tau}_{1.m+1}(s^*|A_2)$, thus sustaining the $F_1$-measurability constraint across the connecting set $A$, and that $\underline{\tau}_{1.m+1}$ remains a sub-strategy that mimics $\tau_2$ (ensuring that $\sum_{s}\underline{\tau}(s|\o) \leq 1$ for every $s$ and $\o$ and the that Inequality \eqref{Inequality of posteriors mimicking} holds).
In conclusion, we constructed a sub-strategy that mimics $\tau_2$ and whose support is $\mathcal{P}^*$, and this concludes the induction.

Let $\underline{\tau}_{1*}$ be the sub-strategy that mimics $\tau_2$ and $\mathbf{P}_{\underline{\tau}_{1*}}(p) >0 $ if and only if $p \in \mathcal{P}^*$.
Assume that $\underline{\tau}_{1*}$ only uses signals in some set $S^*$, that are not used by $\underline{\tau}_{1.0}$ (i.e., $S^* \cap \underline{S} = \phi$).
Define the following sub-strategy
\begin{equation*}
    \underline{\tau}_{2.0}(s|\omega) =
    \begin{cases}
        \underline{\tau}_{1.0}(s|\omega), & \text{for every } (\omega, s) \ \text{s.t. } \underline{\tau}_{1.0}(s|\o)>0, \\
        c \underline{\tau}_{1*}(s|\omega), & \text{for every }  (\omega, s) \ \text{s.t. } \underline{\tau}_{1*}(s|\omega) >0,
    \end{cases}
\end{equation*}
where $c$ is a constant.
Since $\underline{\tau}_{1*}(s|\omega)$ supports only PS posteriors of $\underline{\tau}_{1.0}$, for every state $\omega$
where there exists a PS posterior $p$ of $\underline{\tau}_{1*}(s|\omega)$ such that $\omega\in A_p$,
it follows from Claim 1 that $\sum_{s\in \underline{S}} \underline{\tau}_{1.0}(s|\omega) <1$.
Therefore, by choosing $c$ sufficiently small, we can ensure that
$\sum_{s\in \underline{S}\cup S^*} \underline{\tau}_{2.0}(s|\omega) = \sum_{s\in \underline{S}} \underline{\tau}_{1.0}(s|\omega)  + c \sum_{s\in S^*} \underline{\tau}_{1*}(s|\omega) <1$.
Hence, for the extended strategy $\underline{\tau}_{2.0}(s|\omega)$, we can guarantee that for every $\omega\in \Omega$,
$\sum_{s\in \underline{S}\cup S^*} \underline{\tau}_{2.0}(s|\omega)\leq1$.
We conclude that $\underline{\tau}_{2.0}$ is a sub-strategy that mimics $\tau_2$ and $H(\underline{\tau}_{2.0}) > H(\underline{\tau}_{1.0})$ due to the extension over PS posteriors.
This contradicts the definition of  $\underline{\tau}_{1.0}$ as a mimicking sub-strategy that maximizes $H$.
We can thus conclude that $H(\underline{\tau}_{1.0})=1$, and $\underline{\tau}_{1.0}$ is an $F_1$-measurable strategy that mimics $\tau_2$, as needed.
\hfill
\end{proof}

\subsection{Proof of Proposition
\ref{proposition: balanced} }

\begin{proof} {\bf{iii} } $\Rightarrow$ {\bf{i}}. Suppose that $(\omega_1,\overline{\omega}_1, \omega_2,\overline{\omega}_2 , \dots, \omega_m,\overline{\omega}_m)$ is not $F_2 $-balanced. It means that there is a partition $\{A,B\}$ s.t.\ $\#(A\to B) \not= \#(B\to A)$. Define
\begin{equation*}
f(\o)=
\left\{
               \begin{array}{ll}
                 1, & \hbox{if  }   \o\in A,\\
                 2, & \hbox{if } \o\in B.
               \end{array}
             \right.
\end{equation*}
We obtain, \begin{equation*}
\prod_{i=1}^m
\frac{f(\o_i)}{f(\overline{\omega}_i)}= \left(\frac{1}{2}\right)^{\#(A\to B)} \cdot 2 ^{\#(B\to A)} \not = 1.
\end{equation*} This contradicts {\bf{iii}}.

{\bf{i} } $\Rightarrow$ {\bf{ii}}. Assume {\bf{i} }.
For every $i$, let $D_i= \{\o_j; \o_j \in F_2(\o_i)\} \cup \{\overline{\o}_j; \overline{\o}_j \in F_2(\o_i)\}$ be the set which contains all the states in the loop that share the same information set of $F_2$ as $\o_i$.
Condition {\bf{i} } implies that for every $\o_i$,  the partition $A=D_i$ and $B= (D_i)^c$ satisfies  $\#(A\to B) = \#(B\to A)$.
Note that $|\{\o_j; \o_j\in F_2(\o_i)\}| = \#(A\to B) + \#(A\to A)$,
and $|\{\overline{\o}_j; \overline{\o}_j\in F_2(\o_i)\}| = \#(B\to A) + \#(A\to A)$,
where $\#(A\to A) = |\{i\in\{1,...,m\}; \o_i\in A,\ \overline{\o}_i\in A\}|$. It follows from $\#(A\to B) = \#(B\to A)$ that
\begin{equation}\label{eq: balanced proposition}
|\{\o_j; \o_j\in F_2(\o_i)\}|= |\{\overline{\o}_j; \overline{\o}_j\in F_2(\o_i)\}|
\end{equation}
for every $\o_i$.

Define $J=\{i; \o_i\in F_2(\overline{\o}_i)\}$. We show that the rest of the states are decomposed into $F_2$-loops.
Specifically, we show that if a finite set $S=\{(\o_j, \overline{\omega}_j); \ \overline{\o}_j\notin F_2(\o_j)\}$, not necessarily an $F_1$-loop, satisfies Eq.\ (\ref{eq: balanced proposition}) for every $\o_i\in S$, then it is covered by $F_2$-loops.

When $|S|=2$, Eq.\ (\ref{eq: balanced proposition})  implies that this
is an $F_2$-loop. We now assume the induction hypothesis:
if Eq.\ (\ref{eq: balanced proposition}) is  satisfied for a set $S=\{(\o_j, \overline{\omega}_j)\}$ and for every $\o_i\in S$, and $S$ contains less than or equal to $m$ pairs, then it is covered by $F_2$-loops. We proceed by showing this statement for sets $S$ containing $m+1$ pairs.

We start at an arbitrary pair, say $(\o_1, \overline{\o}_1)$, and show that it belongs to an $F_2$-loop.
Once this $F_2$-loop is formed, the states outside of this loop satisfy Eq.\ (\ref{eq: balanced proposition}) for every $\o_i$ outside of this loop.
By the induction hypothesis, this set is covered by $F_2$-loops.

Due to Eq.\ (\ref{eq: balanced proposition}), there is at least one $\overline{\o}_j$ such that  $\overline{\o}_j \in F_2(\o_1)$.
Consider now the two pairs, $(\o_j, \overline{\o}_j, \o_1, \overline{\o}_1)$.
If this is a loop, Eq.\ (\ref{eq: balanced proposition}) remains true when applied to the states out of this loop.
The induction hypothesis completes the argument.
Otherwise, there is $\overline{\o}_k$ where $k \neq 1,j$, such that $\overline{\o}_k \in F_2(\o_j)$.
Consider now the three pairs, $(\o_k, \overline{\o}_k, \o_j, \overline{\o}_j, \o_1, \overline{\o}_1)$.
If this is an $F_2$-loop, the other states satisfy Eq.\ (\ref{eq: balanced proposition}), and as before, this set is covered by $F_2$-loops.
However, if this is not an $F_2$-loop, Eq.\ (\ref{eq: balanced proposition}) remains true,  we annex another pair and continue this way until we obtain an $F_2$-loop. This loop might cover the entire set, but if not, the remaining states are, by the induction hypothesis, covered by $F_2$-loops. This shows {\bf{ii}}.

{\bf{ii}} $\Rightarrow$ {\bf{iii}}. Let $f:\LB\omega_1,\overline{\omega}_1, \omega_2,\overline{\omega}_2 , \dots, \omega_m,\overline{\omega}_m\RB \to (0, \infty)$ be a positive and $F_2$-measurable function.
Suppose that $I_1,...,I_r$ is a partition of  $\{1,...,m\}$, and for each $t=1,...,r$, the set $\Big((\omega_{i},\overline{\omega}_i)\Big)_{i\in I_t}$ is an $F_2$-loop.
Since, $\Big((\omega_{i},\overline{\omega}_i)\Big)_{i\in I_t}$ is an $F_2$-loop,
$$\prod_{i\in I_t}
\frac{f(\o_i)}{f(\overline{\omega}_i)}=1,
$$
which implies that $$\prod_{i=1}^m \frac{f(\o_i)}{f(\overline{\omega}_i)}=
\prod_{t=1}^r \prod_{i\in I_t} \frac{f(\o_i)}{f(\overline{\omega}_i)}=1.
$$
This proves {\bf{iii}}.
 \end{proof}

\subsection{Proof of Proposition \ref{Proposition - irreducible and informative loops}}

\begin{proof}
Fix an $F_i$-loop $L_i = \Big((\o_j,\overline{\o}_j)\Big)_{j\in I}$ where $I=\{1,2,\dots,m\}$.
Let $C_j$ denote the CKC that contains every pair $(\o_j,\overline{\o}_j)$.

\textbf{Proof for first statement}:
Assume that $L_i$ intersects the same CKC at least twice, so that $C_{l_1}=C_{l_2}$, where $l_1 < l_2$, is such CKC.
Because $L_i$ is a loop, the two pairs $(\o_{l_1},\overline{\o}_{l_1})$ and $(\o_{l_2},\overline{\o}_{l_2})$ that are in this CKC cannot be adjacent in the loop $L_i$, i.e., $l_1\neq l_2\pm 1$.
Define the following sub-loop of $L_i$ by omitting every state from $\overline{\o}_{l_1}$ to $\o_{l_2}$.
Formally, $L_i'=(\o_1, \overline{\o}_1, \dots,\overline{\o}_{l_1-1}, \o_{l_1}, \overline{\o}_{l_2}, \o_{l_2+1},\dots, \o_m,\overline{\o}_m)$.
This is a well-defined sub-loop of $L_i$ (as $\o_{l_1}, \overline{\o}_{l_2} \in C_{l_1}$ while all other parts of the sub-loop match those of $L_i$), which implies that $L_i$ is not irreducible.
Note that the part we truncated from the loop $L_i$ also forms a sub-loop, namely $L_i''=(\o_{l_2}, \overline{\o}_{l_1},\o_{l_1+1}, \overline{\o}_{l_1+1}, \dots, \o_{l_2-1}, \overline{\o}_{l_2-1})$.

\textbf{Proof for second statement}:
Assume, by contradiction, that $L_i$ is irreducible, yet it has a pair of states $(\o_l,\overline{\o}_l)$ such that $\overline{\o}_l\in F_i(\o_l)$.
This implies that $\{\overline{\o}_{l-1},\o_l,\overline{\o}_l,\o_{l+1}\} \subseteq F_i(\o_l)=F_i(\o_{l+1})$.
We can assume that $C_{l-1}\neq C_{l+1}$, otherwise the first statement suggests that $L_i$ is not irreducible.
So, define the following sub-loop of $L_i$ by  $L_i'=\Big((\o_j,\overline{\o}_j)\Big)_{j\in I\setminus\{l\}}$.
Note that $L_i'$ is a well-defined sub-loop, as $C_{l-1}\neq C_{l+1}$ and $\overline{\o}_{l-1} \in F_i(\o_{l+1})$, thus contradicting the irreducible property.

\textbf{Proof for third statement}:
Assume, w.l.o.g., that $F_i(\o_1) \neq F_i(\overline{\o}_1)$.
If $L_i$ intersects the same CKC twice, then we can follow the proof of the first statement, truncate the loop, and take a sub-loop that has an informative pair of states and intersects every CKC at most once.
Thus, w.l.o.g., assume that $L_i$ intersects every CKC at most once.
Denote the set of informative pairs by $I^c=\{j: F_i(\o_j) \neq F_i(\overline{\o}_j)\}$ and define the following ordered sub-loop of $L_i$ by $L_i'=\Big((\o_j,\overline{\o}_j) \Big)_{j\in I^c}$.
In simple terms, $L_i'$ is generated from $L_i$ by truncating all non-informative pairs $(\o_j,\overline{\o}_j)$, where $F_i(\o_j)= F_i(\overline{\o}_j)$, similarly to the process used in the proof of the second statement.
Focusing on $L_i'$, note that: (i) all pairs are pairwise disjoint; (ii) every CKC is crossed at most once; (iii) $\o_{j+1} \in F_i(\overline{\o}_j)$ as we removed only non-informative pairs; and (iv) $\o_j\neq \overline{\o}_j$ are both in the same CKC as in the original loop.
Hence, $L_i'$ is a well-defined loop and an $F_i$-fully-informative sub-loop of $L_i$.

\textbf{Proof of fourth statement:}
If the loop $L_i$ is irreducible, then the statement holds.
Otherwise, it is not irreducible and we will prove by induction on the number of pairs $m$ in $L_1$.
If $m=2$, then $L_i$ is irreducible.
If $m=3$ and $L_i$ is not irreducible, then it has a sub-loop with two pairs.
Assume w.l.o.g.\ that this sub-loop is based on the states $\{\o_1,\overline{\o_1},\o_2, \overline{\o}_2\}$.
It cannot be that $F_i(\overline{\o}_1)=F_i(\overline{\o}_2)$, because that would make $(\o_2,\overline{\o}_2)$ a non-informative pair.
So the sub-loop is $(\o_1,\overline{\o_1},\o_2, \overline{\o}_2)$ such that $F_i(\o_1)=F_i(\overline{\o}_2)$, but $F_i(\o_1)=F_i(\overline{\o}_3)$ and $F_i(\overline{\o}_2)= F_i(\o_3)$, so the pair $(\o_3,\overline{\o}_3)$ is non-informative.

Assume the statement holds for $m=k$ pairs, and consider an $L_i$ loop with $k+1$ pairs.
If the loop intersects the same CKC more than once, we can split is to two sub-loops (as previously done), and use the induction hypothesis for each.
Hence, we can assume that the loop does not intersect the same CKC twice.

Because the loop is not irreducible, there are two states $\o_{i_1}$ and $\overline{\o}_{i_2}$ that are not adjacent in the loop (so $i_1 \geq i_2+2$), yet $F_i(\o_{i_1})= F_i(\overline{\o}_{i_2})$.
The last equality also suggests that $F_i(\overline{\o}_{i_1-1})= F_i({\o}_{i_2+1})$.
If $i_1 = i_2+2$, then there exists only one pair between the two states.
This implies that the pair $(\o_{i_2+1},\overline{\o}_{i_2+1})=(\o_{i_1-1},\overline{\o}_{i_1-1})$ is non-informative, contradicting the fact that $L_i$ is $F_i$-fully-informative.
So we conclude that $i_1 \geq i_2+3$.
Define the following two loops $L_i'=(\o_{i_1},\overline{\o}_{i_1},\dots,\o_{i_2},\overline{\o}_{i_2})$ and $L_i''=({\o}_{i_2+1},\overline{\o}_{i_2+1},\dots,\o_{i_1-1},\overline{\o}_{i_1-1})$, where the ordering of states follows the original loop $L_i$.
These are two well-defined $F_i$-loops with less than $k+1$ pairs each, so the induction hypothesis holds and the result follows.

If $L_i$ does not intersect the same CKC more than once and does not have at least $4$ states in the same partition element, then it is irreducible.

\textbf{Proof of fifth statement:}
If the loop has a non-informative pair $\o_j \in F_i(\overline{\o}_i)$, then it contains $4$ states from the same partition element, so assume that the loop is $F_i$-fully-informative and that it does not intersect the same CKC more than once.
Thus, we need to prove that it has at least $4$ states in the same partition element of $F_i$.

Consider the strict sub-loop $L_i^-$ of $L_i$.
It consists of pairs, taken from the original loop.
Because $L_i$ does not intersect the same CKC more than once, all the pairs of $L_i^-$ are a strict subset of the pairs of $L_i$.
This implies that some pairs were omitted from $L_i$ when generating $L_i^-$, so assume w.l.o.g.\ that the pair $\{\o_1,\overline{\o}_1\}$ is not included in $L_i^-$.
This implies that one pair $\{\o_j,\overline{\o}_j\}$ precedes in $L_i^-$ a different one that it precedes in $L_i$.
That is, $F_i(\overline{\o}_j)=F_i(\o_{j+1})$ according to $L_i$, whereas $F_i(\overline{\o}_j)=F_i(\o_{k})$ where $k\neq j+1$, according to $L_i^-$.
But also $F_i(\o_{k}) = F_i(\overline{\o}_{k-1})$ according  to $L_i$.
Thus, $\{\overline{\o}_j,\o_{j+1},\o_{k},\overline{\o}_{k-1}\}$ are in the same partition element of $L_i$, as stated and the result follows.
    \hfill
\end{proof}

\subsection{Proof of Theorem \ref{Theorem: general case -- necessary condition}}

\begin{proof}
Suppose that  Oracle $1$ dominates Oracle $2$.
If there exists a CKC in which  $F_1$ does not refine $F_2$, Theorem \ref{Theorem - a unique CKC} from Part I (see Appendix \ref{Subsection - a unique CKC}) states that Oracle $1$ does not dominate Oracle $2$ in that CKC.
In other words, there exists $\tau_2$ defined on this CKC, such that for every $\tau_1$, it follows that ${\rm Post}(\tau_1) \nsubseteq {\rm Post}(\tau_2)$.
We extend the definition of $\tau_2$ to the entire state space in an arbitrary way, and still for every $\tau_1$, it follows that ${\rm Post}(\tau_1) \nsubseteq {\rm Post}(\tau_2)$, and we can follow the results of Part I accordingly (specifically, the game of beliefs and Proposition $3$ therein).

We proceed to show that any $F_1$-loop is $F_2$-balanced, which is equivalent to the existence of a cover by loops of $F_2$.
Suppose, to the contrary, that an $F_1$-loop $(\omega_1,\overline{\omega}_1, \omega_2,\overline{\omega}_2 , \dots, \omega_m,\overline{\omega}_m)$ is not $F_2$-balanced.
This means that there is an $F_2$-measurable partition $\{A,B\}$ of these states such that Eq.\ \eqref{eq: balanced} is not satisfied.
We define an $F_2$-measurable signaling function that obtains two signals, $\a$ and $\b$.
Over the states of the loop, let
\begin{equation}\label{eq: D}
\tau_2(\a | \o) =\begin{cases}
      x, & \text{if } \o \in A , \\
       y, & \text{if }  \o \in B,
    \end{cases}
\end{equation}
and $\tau_2(\b | \omega)=1-\tau_2(\a | \omega)$.
On other states, $\tau_2$ is defined arbitrarily.
The numbers $x,y\in (0,1)$ are chosen so that $\frac{\ln{x}-\ln{y}}{\ln{(1-x)}-\ln{(1-y)}}$ is irrational.

\heading{Claim 1:} If  ${\rm Post}(\tau_1) \subseteq {\rm Post}(\tau_2)$, then any signal of $\tau_1$ induces the same posteriors as $\a$ does or as $\b$ does in every CKC.

\heading{Claim 2:} For any signal $s$ of $\tau_1$ and for any $i$,
$\frac{\tau_1(s|\o_i)}{\tau_1(s|\overline{\omega_i})} \in \{\frac{x}{y},\frac{1-x}{1-y},\frac{y}{x},  \frac{1-y}{1-x}\}$ .
Therefore,
$$\prod_{i=1}^m  \frac{\tau_1(s|\o_i)}{\tau_1(s|\overline{\omega}_i)}=
\left(\frac{x}{y}\right)^{\l_1}\cdot
\left(\frac{1-x}{1-y}\right)^{\l_2}\cdot
\left(\frac{y}{x}\right)^{k_1}\cdot
\left(\frac{1-y}{1-x}\right)^{k_2},$$
where $\l_1+\l_2= |\{i; \o_i\in A \ {\rm{and}} \ \overline{\omega}_i\in B\}|$ and
$k_1+k_2= |\{i; \o_i\in B \ {\rm{and}} \ \overline{\omega}_i\in A\}|$.

\heading{Claim 3:} For any signal $s$ of $\tau_1$, $\prod_{i=1}^m  \frac{\tau_1(s|\o_i)}{\tau_1(s|\overline{\omega}_i)}=1$.

We therefore obtain $(\frac{x}{y})^{\l_1}(\frac{1-x}{1-y})^{\l_2}(\frac{y}{x})^{k_1}(\frac{1-y}{1-x})^{k_2}=1$. We conclude that there are whole numbers, say $\l=\l_1-k_1$ and $ k=k_2-\l_2$ such that $(\frac{x}{y})^{\l }=(\frac{1-x}{1-y})^{k}$.
Since $\frac{\ln{x}-\ln{y}}{\ln{(1-x)}-\ln{(1-y)}}=\frac{\ln{\frac{x}{y}}}{\ln{\frac{1-x}{1-y}}}$ is irrational, $\l=k=0$, implying that Eq.\ (\ref{eq: balanced}) is satisfied.
This is a contradiction, so every $F_1$-loop is $F_2$-balanced.

Moving on to the third part of the theorem, fix an irreducible $F_1$-loop $L_1$, and consider an irreducible cover by a unique $F_2$-loop $L_2$, i.e., $L_2$ covers $L_1$ and both are irreducible w.r.t.\ the relevant partition.
Note that if $L_2$ is also order-preserving, it implies that it \emph{matches} $L_1$.

Assume, by contradiction, that $L_2$ is not order-preserving and the two loops do not match one another.
Denote $L_1 = (\omega_1, \overline{\omega}_1, \dots, \omega_m, \overline{\omega}_m)$ and  $L_2 = (\omega_1,\overline{\omega}_1, \omega_{i_2}, \overline{\omega}_{i_2}, \dots, \omega_{i_m}, \overline{\omega}_{i_m})$.
Thus, there exist indices $k>j>1$ such that $\o_k$ precedes $\o_j$ in $L_2$.
In simple terms, it implies that though $L_2$ consists of the same pairs as $L_1$, the ordering of pairs throughout the two loops differs, as suggested in Footnote \ref{Footnote - order of subloop}.

Since the two loops are irreducible, it follows from Proposition \ref{Proposition - irreducible and informative loops} that they intersect every CKC at most once and that both are fully-informative.
Moreover, for every state $\o$ in every loop $L_i$, every set $F_i(\o)$ contains two states from the loop $L_i$ (otherwise, the loop is not irreducible).
So, one can define an $F_i$-measurable function $\tau_i$ such that $\tau_i(s|\o_l) = \tau_i(s|\overline{\o}_{l-1}) \neq \tau_i(s|\o_{l'})$ for every $\o_l \neq \o_{l'}$ in the loop.

To simplify the exposition, partition the states of $L_2$ into three disjoint sets: the set $A^2_1=\{\overline{\o}_1,\dots,\o_{k}\}$ contains all the states of $L_2$ from $\overline{\o}_1$ till $\o_k$ (following the order of $L_2$), $A^2_k=\{\overline{\o}_k,\dots,\o_{j}\}$ contains all the states of $L_2$ from $\overline{\o}_k$ till $\o_j$, and $A^2_j=\{\overline{\o}_j,\dots,\o_{1}\}$ which contains all remaining states of $L_2$.
Follow a similar process with $L_1$, so that $A^1_1=\{\overline{\o}_1,\dots,\o_{j}\}$ contains all the states of $L_1$ from $\overline{\o}_1$ till $\o_j$ (following the order of $L_1$), $A^1_j=\{\overline{\o}_j,\dots,\o_{k}\}$ contains all the states of $L_1$ from $\overline{\o}_j$ till $\o_k$, and $A^1_k=\{\overline{\o}_k,\dots,\o_{1}\}$ which contains all remaining states of $L_1$.

Denote by $C_l$ the CKC of the pair $(\omega_l,\overline{\omega}_l)$.
Fix two distinct signals $s_1$ and $s_2$, and define the signaling function $\tau_2$ as follows:
\begin{equation*}
    \tau_2(s_1|\omega) = 1-\tau_2(s_2|\omega) = \begin{cases}
       p_1, & \text{if } \omega \in A^2_1=\{\overline{\o}_1,\dots,\o_{k}\}, \\
       p_2, & \text{if } \omega \in A^2_k=\{\overline{\o}_k,\dots,\o_{j}\}, \\
       p_3, & \text{if } \omega \in A^2_j=\{\overline{\o}_j,\dots,\o_{1}\}, \\
       p_4, & \text{if } \omega \in \Omega \setminus \bigcup_{i=1,j,k} A^2_i, \\
    \end{cases}
\end{equation*}
where the probabilities $\{p_1,p_2,p_3,p_4\}$ are chosen as in the strategy defined in Equation \eqref{Equation - fully support tau_2 on m elements}.
Because the loop is irreducible, intersects every CKC at most once and $F_2$-fully-informative, $\tau_2$ is a well-defined $F_2$-measurable function.

The result of Lemma \ref{Lemma - proportional signals with m states} from Part I (see Appendix \ref{Subsection - proportional signals with m states}) holds in every CKC of the loop (though with different probabilities).
So given a CKC $C_l$, if there exists $\tau_1$ such that ${\rm Post}(\tau_1) \subseteq {\rm Post}(\tau_2)$, then for every signal $t \in {\rm Supp}(\tau_1)$ there exists a signal $s \in \{s_1,s_2\}$ and a constant $c>0$ such that $\tau_1(t|\omega)=c\tau_2(s|\omega)$ for every $\omega \in C_l$.
Therefore, in every CKC $C_l$ and for every signal $t$, there exists a signal $s$ such that $\tfrac{\tau_2(s|\o_l)}{\tau_2(s|\overline{\o}_l)} = \tfrac{\tau_1(t|\o_l)}{\tau_1(t|\overline{\o}_l)}$.
Fix such a strategy $\tau_1$.

Notice that in every CKC $C_l \neq C_1,C_j,C_k$ and for every signal $s\in \{s_1,s_2\}$, we get $\tau_2(s|\o_l) = \tau_2(s|\overline{\o}_l)$.
Thus, $\tfrac{\tau_1(t|\o_l)}{\tau_1(t|\overline{\o}_l)}=1$ for every $t$ and every $l \neq i,j,k$.
This implies that for every feasible signal $t$ restricted to the loop $L_1$,
\begin{equation*}
    \tau_1(t|\omega) =
    \begin{cases}
       a_t, & \text{if } \omega \in A^1_1=\{\overline{\o}_1,\dots,\o_{j}\}, \\
       b_t, & \text{if } \omega \in A^1_j=\{\overline{\o}_j,\dots,\o_{k}\}, \\
       c_t, & \text{if } \omega \in A^1_k=\{\overline{\o}_k,\dots,\o_{1}\},
    \end{cases}
\end{equation*}
where $a_t, b_t, c_t \in (0,1]$.
Evidently, the parameters $a_t,b_t$, and $c_t$ can vary across the feasible signals.

In addition, Lemma \ref{Lemma - proportional signals with m states} from Part I (see Appendix \ref{Subsection - proportional signals with m states}) states that in every CKC, $\tau_1(t|\omega)$ is proportional to $\tau_2(s_i|\omega)$ for some signal $s_i \in \{s_1,s_2\}$.
This yields the following constraints:
$$
\frac{\tau_1(t|\omega_1)}{\tau_1(t|\overline{\omega}_1)} = \frac{c_t}{a_t} =  \frac{\tau_2(s_i|\omega_1)}{\tau_2(s_i|\overline{\omega}_1)} \in \Big\{\frac{p_3}{p_1}, \frac{1-p_3}{1-p_1} \Big\},
$$
$$
\frac{\tau_1(t|\omega_j)}{\tau_1(t|\overline{\omega}_j)} = \frac{a_t}{b_t} =  \frac{\tau_2(s_i|\omega_j)}{\tau_2(s_i|\overline{\omega}_j)} \in \Big\{\frac{p_2}{p_3}, \frac{1-p_2}{1-p_3} \Big\},
$$
$$
\frac{\tau_1(t|\omega_k)}{\tau_1(t|\overline{\omega}_k)} = \frac{b_t}{c_t} =  \frac{\tau_2(s_i|\omega_k)}{\tau_2(s_i|\overline{\omega}_k)} \in \Big\{\frac{p_1}{p_2}, \frac{1-p_1}{1-p_2} \Big\}.
$$
Because the two loops cover one another and specifically because $L_2$ is $F_1$-covered, Proposition \ref{proposition: balanced} states that $\prod_{l=1}^m \frac{\tau_1(t|\omega_{i_l})}{\tau_1(t|\overline{\omega}_{i_l})}=1$, which leaves only two possibilities for the ratios $\{\tfrac{c_t}{a_t}, \tfrac{a_t}{b_t}, \tfrac{b_t}{c_t}\}$ above: either they equal $\{\tfrac{p_3}{p_1}, \tfrac{p_2}{p_3}, \tfrac{p_1}{p_2}\}$ respectively, or $\{\tfrac{1-p_3}{1-p_1}, \tfrac{1-p_2}{1-p_3}, \tfrac{1-p_1}{1-p_2}\}$.
This follows from the uniqueness of the ratios, as stated in Lemma \ref{Lemma - proportional signals with m states} from Part I (see Appendix \ref{Subsection - proportional signals with m states}).
Note that this must hold for every feasible signal $t$ of $\tau_1$ across the loop.

\begin{figure}[th!]
\centering
\begin{tabular}{c|c|c|}
    $\tau_1(t|\o)$ & $t_1$ & $t_2$ \\
\hline
$\o_1$ & $\lambda_1 c_1$ & $\lambda_2 c_2$ \\
\hline
$\overline{\o}_1$ & $\lambda_1 a_1$ & $\lambda_2 a_2$ \\
\hline
$\o_j$ &  $\lambda_1 a_1$ &  $\lambda_2 a_2$ \\
\hline
$\overline{\o}_j$ &  $\lambda_1 b_1$  &  $\lambda_2 b_2$  \\
\hline
$\o_k$ &  $\lambda_1 b_1$   &  $\lambda_2 b_2$   \\
\hline
$\overline{\o}_k$ &$\lambda_1 c_1$  &  $\lambda_2 c_2$ \\
\hline
\end{tabular}
\caption{ \footnotesize The structure of $\tau_1$ restricted to the states $\{\omega_1, \overline{\omega}_1, \omega_j, \overline{\omega}_j, \omega_k, \overline{\omega}_k \}$, where $\frac{c_1}{a_1} = \frac{p_3}{p_1}$, $\frac{b_1}{c_1} = \frac{p_1}{p_2}$, $\frac{c_2}{a_2} = \frac{1-p_3}{1-p_1}$ and  $\frac{b_2}{c_2} = \frac{1-p_1}{1-p_2}$ and $\lambda_1,\lambda_2 >0$.}
\label{fig:first table of tau1 in equivalent theorem}
\end{figure}

Thus, if we focus on the states $\{\omega_1, \overline{\omega}_1, \omega_j, \overline{\omega}_j, \omega_k, \overline{\omega}_k \}$ and group together all signals $t$ with the same distribution on these states, then for some positive constants $\lambda_1,\lambda_2>0$ we get the strategy defined in Figure \ref{fig:first table of tau1 in equivalent theorem}.
Plugging in the relevant ratios yields the probabilities given in Figure \ref{fig:second table of tau1 in equivalent theorem}.

\begin{figure}[th!]
\centering
\begin{tabular}{c|c|c|}
    $\tau_1(t|\o)$ & $t_1$ & $t_2$ \\
\hline
$\o_1$ & $\lambda_1 c_1$ & $\lambda_2 c_2$ \\
\hline
$\overline{\o}_1$ & $\lambda_1 c_1 \frac{p_1}{p_3}$ & $\lambda_2 c_2 \frac{1-p_1}{1-p_3}$ \\
\hline
$\o_j$ &  $\lambda_1 c_1 \frac{p_1}{p_3}$  &  $\lambda_2 c_2 \frac{1-p_1}{1-p_3}$  \\
\hline
$\overline{\o}_j$ &  $\lambda_1 c_1 \frac{p_1}{p_2}$  &  $\lambda_2 c_2 \frac{1-p_1}{1-p_2}$   \\
\hline
\end{tabular}
\caption{ \footnotesize The structure of $\tau_1$ restricted to the states $\{\omega_1, \overline{\omega}_1, \omega_j, \overline{\omega}_j\}$,  where probabilities are presented in terms of $c_1, c_2, \lambda_1$ and $\lambda_2$.}
\label{fig:second table of tau1 in equivalent theorem}
\end{figure}

Recall that the rows must sum to $1$, so that $\tau_1$ is a well-defined strategy.
So, we get the following system of linear equations, in which $(x,y)=(\lambda_1c_1,\lambda_2c_2)$ and:
\begin{eqnarray*}
    x+y & = & 1, \\
    \frac{p_1}{p_3}x+\frac{1-p_1}{1-p_3}y & = & 1, \\
    \frac{p_1}{p_2}x+\frac{1-p_1}{1-p_2}y & = & 1,
\end{eqnarray*}
which does not have a solution since $p_1,\ p_2,\ p_3$ are required to be distinct.
Thus, we conclude that the loops must sustain the same ordering of pairs, and therefore coincide as needed.
This concludes the third and final part of the theorem.
\hfill
\end{proof}

\subsection{Proof of Theorem \ref{Theorem: NI leads to dominance}}

\begin{proof}
We first define an auxiliary set $\overline{\Omega}$, which groups together states that are in the same partition element of $F_2$ within CKCs.
Formally, define the set $\overline{\Omega}$ such that $\eta(\omega') \in \overline{\Omega}$ if and only if $\eta(\omega') = \{\omega \in \Omega : \  \omega,\omega' \in C_j, \ F_2(\omega) =F_2(\omega')  \}$.
Accordingly, define the partition $\overline{F_2}$ to be discrete in every CKC, such that $\overline{F_2}(\eta(\omega)) = \overline{F_2}(\eta(\omega'))$ if and only if $F_2(\omega) = F_2(\omega')$.
Note that $\overline{F_2}$ is essentially a projection of $F_2$ onto $\overline{\Omega}$.
In addition, $\overline{F_1}$ is defined as follows: (i) discrete in every CKC, similarly to $\overline{F_2}$; (ii) $\overline{F_1}(\eta(\omega)) = \overline{F_1}(\eta(\omega'))$ if $\omega$ and $\omega'$ are not in the same CKC, and there exist $\overline{\omega} \in \eta(\omega)$ and $\overline{\omega'} \in \eta(\omega')$ such that $F_1(\overline{\omega}) =F_1(\overline{\omega'})$; and (iii) $\overline{F_1}$ forms a partition (i.e., given (i) and (ii), if two elements of $\overline{F_1}$ contain the same state $\eta(\omega)$, they are unified into one element).

We now prove that $\overline{F_1}=\overline{F_2}$ in every CKC and that there are no $\overline{F_1}$-loops.
Thus, by Theorem \ref{Theorem - no loop characterization}, any $\overline{F_2}$-measurable strategy $\overline{\tau_2}$ (which, extended to $\Omega$, is also $F_2$-measurable) can be imitated by an $\overline{F_1}$-measurable strategy $\overline{\tau_1}$.

\noindent \textbf{Step 1: $\overline{F_1}=\overline{F_2}$ in every CKC.}

By definition, $\overline{F_2}$ refines $\overline{F_1}$, so we need to prove that $\overline{F_1}$ also refines $\overline{F_2}$ in every CKC.
Assume, by contradiction, that $\overline{F_1}(\eta(\omega)) = \overline{F_1}(\eta(\omega'))$ where $\omega$ and $\omega'$ are in the same CKC, whereas $\overline{F_2}(\eta(\omega)) \neq \overline{F_2}(\eta(\omega'))$.
This suggests that $F_2(\omega) \neq F_2(\omega')$, which implies that $F_1(\omega) \neq F_1(\omega')$.
According to the construction of $\overline{F_1}$, we conclude that the equality $\overline{F_1}(\eta(\omega)) = \overline{F_1}(\eta(\omega'))$ followed from the partition-formation stage described in (iii) above, through at least one other CKC.
Thus, there exists an $F_1$-loop which connects a state in $\eta(\omega)$ with a state in $\eta(\omega')$.
Without loss of generality, assume these states are $\omega$ and $\omega'$.
Because every $F_1$-loop is $F_2$-non-informative, it follows that $F_2(\omega) = F_2(\omega')$, a contradiction.

\noindent \textbf{Step 2: There are no $\overline{F_1}$-loops.}

An $\overline{F_1}$-loop implies that an $F_1$-loop exists.
By construction, all $\Omega$ states in every CKC are $F_2$-equivalent (i.e., grouped together according to $F_2$).
Because every $F_1$-loop is $F_2$-non-informative, it implies that the loop consists of only one $\overline{\Omega}$ state in every CKC, and not two.
This contradicts the definition of a loop.

\noindent \textbf{Step 3: $\overline{F_1}$ can mimic $\overline{F_2}$.}

Fix a strategy $\tau_2$, and let $\overline{\tau_2}$ be the projected strategy on $\overline{\Omega}$.
Because $\overline{F_1}=\overline{F_2}$ in every CKC and there are no $\overline{F_1}$-loops, there exists an $\overline{F_1}$-measurable  strategy $\overline{\tau_1}$ that imitates $\overline{\tau_2}$.
Therefore, one can lift $\overline{\tau_1}$ to $\Omega$ to create $\tau_1$, whose projection onto $\overline{\Omega}$ matches $\overline{\tau_1}$.
Thus, the strategy $\tau_1$ imitates $\tau_2$, as needed.
    \hfill
\end{proof}

\subsection{Proof of Proposition \ref{Proposition: two CKCs}}

\begin{proof}
Denote the two CKCs by $C_1$ and $C_2$.
One part of the statement follows directly from Theorem \ref{Theorem: general case -- necessary condition}, so assume that $F_1$ refines $F_2$ in every CKC and any $F_1$-loop is $F_2$-balanced.
If there are no $F_1$-loops, then the result follows from Theorem \ref{Theorem - no loop characterization}, so assume there exists at least one $F_1$-loop, and every such loop is $F_2$-balanced.

Take any $F_1$-loop $(\omega_1,\overline{\omega}_1, \omega_2,\overline{\omega}_2)$ with four states.
We argue that either it is also an $F_2$-loop or it is $F_2$-non-informative.
Otherwise, we can assume (without loss of generality) that $F_2(\omega_1) \neq F_2(\overline{\omega}_i)$, for every $i=1,2$.
So, there are only two possibilities left: either $F_2(\omega_1) = F_2(\omega_2)$ or $F_2(\omega_1) \neq F_2(\omega_2)$.
If $F_2(\omega_1) = F_2(\omega_2)$, then there exists an $F_2$-measurable partition of the four states such that $A=\{\omega_1,\omega_2\}$ and $B=\{\overline{\omega}_1,\overline{\omega}_2\}$, which is not balanced.
Otherwise, there exists another non-balanced $F_2$-measurable partition of the form $A=\{\omega_1\}$ and $B=\{\overline{\omega}_1,\omega_2,\overline{\omega}_2\}$.
In any case, we get a contradiction.

The proof now splits into two cases: either there exists an $F_1$-loop $(\omega_1,\overline{\omega}_1, \omega_2,\overline{\omega}_2)$ and an index $i$ such that $ F_2(\omega_i) \neq F_2(\overline{\omega_i})$, or every such loop is $F_2$-non-informative.
If indeed every such loop is $F_2$-non-informative, Theorem \ref{Theorem: NI leads to dominance} states that Oracle $1$ dominates Oracle $2$, so we need only focus on the former.

Assume that there exists an $F_1$-loop $(\omega_1,\overline{\omega}_1, \omega_2,\overline{\omega}_2)$ and an index $i$ such that $F_2(\omega_i) \neq F_2(\overline{\omega}_i)$.
Denote this couple by $\{\omega_1,\overline{\omega}_1\} \subseteq C_1$.
The previous conclusion implies that it is also an $F_2$-loop.
We claim that, under these conditions, every $\tau_2$ is $F_1$-measurable.
Note that $F_1$ refines $F_2$ in every CKC, so we need to verify that for every $(\omega,\overline{\omega})\in C_1\times C_2$ such that $F_1(\omega)= F_1(\overline{\omega})$, it follows that $F_2(\omega)= F_2(\overline{\omega})$.

Take $(\omega,\overline{\omega})\in C_1\times C_2$ such that $F_1(\omega)= F_1(\overline{\omega})$.
If  $\omega = \omega_1$ or $\omega = \overline{\omega}_1$, then $(\omega,\overline{\omega})$ are part of the previously stated $F_2$-loop, so $F_2(\omega)= F_2(\overline{\omega})$.
Otherwise, we can construct two new $F_1$-loops $(\omega,\overline{\omega}, \omega_1,\overline{\omega}_2)$ and $(\omega,\overline{\omega}, \omega_2,\overline{\omega}_1)$.
Because $F_2(\omega_1) \neq F_2(\overline{\omega}_1)$, either $F_2(\omega) \neq F_2(\omega_1)$ or $F_2(\omega) \neq F_2(\overline{\omega}_1)$.
The previous conclusion again implies that $(\omega, \overline{\omega})$ are a apart of an $F_2$-loop, so $F_2(\omega)= F_2(\overline{\omega})$, as needed.
\hfill
\end{proof}

\subsection{Proof of Theorem \ref{Theorem: Equivalent oracles}}

\begin{proof}
We start by assuming that $F_1$ and $F_2$ are equivalent.
According to Theorem \ref{Theorem: general case -- necessary condition}, every $F_i$ refines $F_{-i}$ in every CKC, and every $F_i$-loop is covered by $F_{-i}$-loops.
Fix an irreducible $F_i$-loop with at least $6$ states, denoted $L_i$, and consider a cover by $F_{-i}$-loops.
There are two possibilities: either the cover constitutes a single loop, or else.
If the cover contains a shorter loop, say $L_{-i}'$, then that loop is not $F_i$-covered because $L_i$ is irreducible, and this contradicts Theorem \ref{Theorem: general case -- necessary condition}.
Moreover, the cover cannot have non-informative pairs where $F_{-i}(\o_i)=F_{-i}(\overline{\o}_i)$, because the two partitions match one another in every CKC and $L_i$ is irreducible.
So, the cover consists of a single irreducible $F_{-i}$-loop, and Theorem \ref{Theorem: general case -- necessary condition} states that it is order-preserving.
Thus, $L_i$ and $L_{-i}$ coincide as stated.

Moving to the other direction, assume that $F_i$ refines $F_{-i}$ in every CKC, that any $F_i$-loop has a cover of $F_{-i}$-loops, and every irreducible $F_i$-loop with at least $6$ states is an irreducible $F_{-i}$-loop.
Let us prove that Oracle $1$ dominates Oracle $2$ (and the reverse dominance follows symmetrically).

We start with two simple observations.
First, in case $F_1$ has no loops, then the statement follows from previous results, so assume $F_1$ has loops.
Second, we say that two CKCs $C_1$ and $C_2$ are connected if there exist $\omega_1 \in C_1$ and $\omega_2 \in C_2$ such that $F_1(\omega_1)=F_1(\omega_2)$.
If there exists a CKC $C$ which is not connected to any other CKC (i.e., for every $\o\in C$, the partition element $F_1(\omega) \subseteq C$), then Oracle $1$ dominates Oracle $2$ conditional on that CKC and independently of all other CKCs.
Thus, without loss of generality, we can assume that all CKCs are connected, either directly or sequentially.

For this part, we will need to define the notion of \emph{type-2 irreducible} loops, which are fully-informative loops that do not have four states in the same information set of the relevant $F_i$.

\begin{definition}
    Let $L_i$ be an $F_i$-loop.
    We say that the loop is \emph{type-2 irreducible} if it does not have four states in the same information set \emph{(}i.e., partition element\emph{)} of $F_i$.
\end{definition}

We shall use this notion of type-2 irreducible $F_1$-loops as building blocks upon which every $F_2$-measurable $\tau_2$ is also $F_1$-measurable.
For that purpose, we start by proving in the following Claim \ref{Claim - Every type-2 irreducible $F_1$-loop is an $F_2$-loop} that every type-2 irreducible $F_1$-loop is also an $F_2$-loop.
Next, we will extend this measurability result to every set of type-2 irreducible $F_1$-loops that intersect the same CKCs, and finally extend it to all CKCs that these loops intersect.
This sets of CKCs, to be later defined as \emph{clusters}, will be the basic sets upon which every $F_2$-measurable strategy is also $F_1$-measurable.

\begin{claim} \label{Claim - Every type-2 irreducible $F_1$-loop is an $F_2$-loop}
Every type-2 irreducible $F_1$-loop $L_1$ is an $F_2$-loop.
\end{claim}

\begin{proof}
If $L_1$ is irreducible, then it is also an irreducible $F_2$-loop, and the result holds.
Thus assume that $L_1$ is not irreducible.
Using the fifth result in Proposition \ref{Proposition - irreducible and informative loops}, we deduce that $L_1$ intersects the same CKC more than once.
Using the proof of the first result in Proposition \ref{Proposition - irreducible and informative loops}, we can decompose $L_1$ into two disjoint strict sub-loops of $F_1$.
This can be done repeatedly, so that $L_1$ is decomposed into sub-loops that do not intersect the same CKC more than once.
This implies that every such loop is type-2 irreducible.
Thus, every such sub-loop is irreducible, and so it is also an $F_2$-loop.

Note that the decomposition process occurs \emph{within} every relevant CKC $C$ and that $F_1|_C=F_2|_{C}$.
That is, once there are two pairs of the same loop within the same CKC, we can decompose the loop into two disjoint loops by rearranging these four states.
So, one can reverse the process and recompose the sub-loops of $F_2$ to regenerate the original loop $L_1$, which is now also an $F_2$-loop, as needed.
\hfill
\end{proof}

Once we dealt with individual type-2 irreducible loops, we move to loops that intersect the same CKC.
For that purpose, we need to prove the following supporting, general Claim \ref{Claim - decomposition of FI loops to type-2 loop} which states that every  $F_i$-fully-informative loop $L_i$ can be decomposed to type-2 irreducible  $F_i$-loops.

\begin{claim} \label{Claim - decomposition of FI loops to type-2 loop}
Every $F_i$-fully-informative loop $L_i$ that is not type-2 irreducible can be decomposed to type-2 irreducible  $F_i$-loops.
\end{claim}

\begin{proof}
The proof is done by induction on the number of pairs $m$ in $L_i$.
If $m=2$, then it is irreducible, as needed.
Assume that the statement holds for $m=k$, and consider a loop with $k+1$ pairs.
If it is not type-2 irreducible, then it has four different states $\{\overline{\o}_j, \o_{j+1}, \overline{\o}_l, \o_{l+1}\}$ in the same information set of $F_i$, where $l>j+1$ and $l+1<j$ so that the two pairs are not adjacent in the original loop $L_i$ (otherwise, the loop has a non-informative pair).
Note that an additional connection may exist, but in any case $\o_{j+1}$ is in the same partition element as $\overline{\o}_j$, and the same holds for $\overline{\o}_l$ and $\o_{l+1}$.
Consider the loops $(\o_j,\overline{\o}_j,\o_{l+1},\overline{\o}_{l+1},\o_{l+2},\overline{\o}_{l+2},\dots,\o_{j-1},\overline{\o}_{j-1})$ and $(\o_{l},\overline{\o}_{l},\o_{j+1},\overline{\o}_{j+1},\o_{j+2},\overline{\o}_{j+2},\dots,\o_{l-1},\overline{\o}_{l-1})$.
The two sub-loops are based on the original loop, other than the first pair, see Figure \ref{Figure- FI loop not type-2 irreducible}

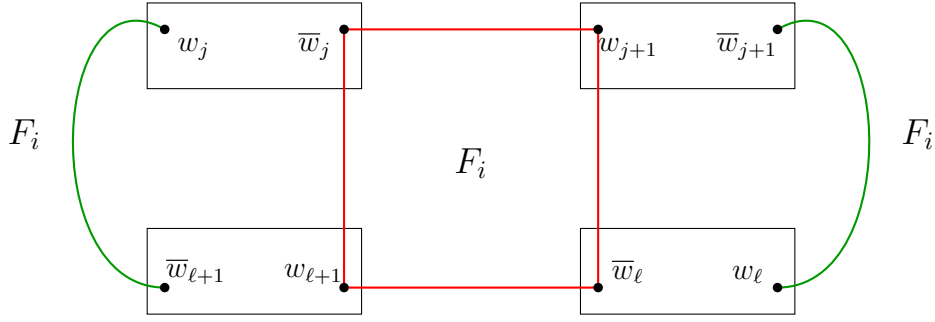
\begin{figure}[H]
\centering
\begin{tikzpicture}[scale=1.2, every node/.style={scale=0.95}]
  \node[draw, rectangle, minimum width=3cm, minimum height=1.2cm] (A) at (0,2.5) {$w_j \quad \quad \quad \overline{w}_j$};
  \node[draw, rectangle, minimum width=3cm, minimum height=1.2cm] (B) at (4.8,2.5) {$w_{j+1} \quad \quad \overline{w}_{j+1}$};
  \node[draw, rectangle, minimum width=3cm, minimum height=1.2cm] (C) at (0,0) {$\overline{w}_{\ell+1} \quad \quad  w_{\ell+1}$};
  \node[draw, rectangle, minimum width=3cm, minimum height=1.2cm] (D) at (4.8,0) {$\overline{w}_{\ell} \quad \quad \quad w_{\ell}$};

  \coordinate (dot_wj)      at ($(A.north west) + (0.2,-0.3)$);
  \coordinate (dot_wjbar)   at ($(A.north east) + (-0.2,-0.3)$);
  \coordinate (dot_wj1)     at ($(B.north west) + (0.2,-0.3)$);
  \coordinate (dot_wj1bar)  at ($(B.north east) + (-0.2,-0.3)$);

  \coordinate (dot_wl1bar)  at ($(C.south west) + (0.2,0.3)$);
  \coordinate (dot_wl1)     at ($(C.south east) + (-0.2,0.3)$);
  \coordinate (dot_wlbar)   at ($(D.south west) + (0.2,0.3)$);
  \coordinate (dot_wl)      at ($(D.south east) + (-0.2,0.3)$);

  \draw[red, thick] (dot_wjbar) -- (dot_wj1);
  \draw[red, thick] (dot_wj1) -- (dot_wlbar);
  \draw[red, thick] (dot_wlbar) -- (dot_wl1);
  \draw[red, thick] (dot_wl1) -- (dot_wjbar);
  \node[black] at (2.4,1.2) {\large $F_i$};

  \draw[green!60!black, thick, .]
    (dot_wj) to[out=150, in=180, looseness=1.3] node[left,xshift=-0.3cm] {\large $F_i$} (dot_wl1bar);
  \draw[green!60!black, thick, .]
    (dot_wj1bar) to[out=30, in=0, looseness=1.3] node[right,xshift=0.3cm] {\large $F_i$} (dot_wl);

  \foreach \pt in {dot_wj, dot_wjbar, dot_wj1, dot_wj1bar, dot_wl1bar, dot_wl1, dot_wlbar, dot_wl}
    \fill (\pt) circle (1.5pt);

\end{tikzpicture}
\caption{A fully-informative loop that is not type-2 irreducible, with four states in the same information set of $F_i$. The red rectangle denotes the same partition element of $F_i$, and the green edges denote the additional states of the original loop.}\label{Figure- FI loop not type-2 irreducible}
\end{figure}

Each of these sub-loops is $F_i$-fully-informative, and has strictly less than $k$ pairs.
Thus, the induction hypothesis holds, and they are either type-2 irreducible, or can be separately decomposed to type-2 irreducible loops, so the result follows.

Note that even without the induction hypothesis, we can repeat the decomposition process, so that all the connections of the original loop that are based on information sets of $F_i$ with no more than two states (in the loop) are kept in one of the sub-loops.
    \hfill
\end{proof}

Using Claim \ref{Claim - decomposition of FI loops to type-2 loop}, we now prove in the following Claim \ref{Claim - type-2 irreducible $F_1$-loop with joint CKC}, that every $F_2$-measurable strategy on two type-2 irreducible $F_1$-loops with a joint CKC (i.e., pass through the same CKC) is $F_1$-measurable.

\begin{claim} \label{Claim - type-2 irreducible $F_1$-loop with joint CKC}
Fix two type-2 irreducible $F_1$-loops $L_1$ and $L_1'$ that share at least one \emph{CKC}.
Then, every $\tau_2|_{L_1\cup L_1'}$ is $F_1$-measurable.
\end{claim}

\begin{proof}
Fix two type-2 irreducible $F_1$-loop $L_1$ and $L_1'$, and assume that they share at least one CKC.
Denote $L_1=(\omega_1, \overline{\omega}_1, \omega_2, \overline{\omega}_2, \dots, \omega_m, \overline{\omega}_m)$ and $L_1'=(\omega_1', \overline{\omega}_1', \omega_2', \overline{\omega}_2', \dots, \omega_{m'}', \overline{\omega}_{m'}')$.
Assume, by contradiction, that there exists a strategy $\tau_2|_{L_1\cup L_1'}$ which is not $F_1$-measurable.
As already proven, each of these loops is also an $F_2$-loop, so the measurability constraint implies that there exist $\o \in L_1$ and $\o'\in L_1'$ such that $F_2(\o) \neq F_2(\o')$ whereas $F_1(\o) = F_1(\o')$.
Because $F_1$ and $F_2$ match one another in every CKC, this suggests that $\o$ and $\o'$ are in two different CKCs.
Denote a shared CKC by $C_j$ in which there are the pairs $(\omega_j, \overline{\omega}_j)$ and $(\omega_j', \overline{\omega}_j')$ taken from $L_1$ and $L_1'$ respectively.
Note that the two pairs may coincide, as well as contain one of the states $\o$ and  $\o'$, but not both (because the two are in different CKCs).
See Figure \ref{Figure - Two type-2 irreducible loops of $F_1$ that share at least one CKC}

\begin{figure}[H]
\centering
\begin{tikzpicture}[scale=1.2, every node/.style={scale=0.95}]

  \draw[blue, thick] (2,0.8) rectangle (4.4,-0.8);
  \node[blue] at (3.2,1.1) {\small $C_j$};

  \node at (2.6,0.55) {$\overline{w}_j$};
  \fill (2.6,0.25) circle (1.5pt);
  \coordinate (dot_wjbar) at (2.6,0.25);

  \node at (3.8,0.55) {$w_j$};
  \fill (3.8,0.25) circle (1.5pt);
  \coordinate (dot_wj) at (3.8,0.25);

  \node at (2.6,-0.55) {$\overline{w}'_j$};
  \fill (2.6,-0.25) circle (1.5pt);
  \coordinate (dot_wjbarp) at (2.6,-0.25);

  \node at (3.8,-0.55) {$w'_j$};
  \fill (3.8,-0.25) circle (1.5pt);
  \coordinate (dot_wjp) at (3.8,-0.25);

  \draw[thick] (dot_wjbar)
    to[out=155, in=25, looseness=12] (dot_wj);
  \node at (3.2,1.85) {\small $F_1$};

  \draw[thick] (dot_wjbarp)
    to[out=-155, in=-25, looseness=12] (dot_wjp);
  \node at (3.2,-1.85) {\small $F_1$};

  \node[purple] at (0.9,1.4) {$w_1$};
  \fill (1.25,1.4) circle (1.5pt);

  \node[purple] at (0.9,-1.4) {$\overline{w}'_1$};
  \fill (1.25,-1.4) circle (1.5pt);

  \draw[purple, dashed, thick] (1.25,1.4) -- (1.25,-1.4);
  \node[purple] at (1,0) {\small $F_1$};

\end{tikzpicture}
\caption{Two type-2 irreducible loops of $F_1$ that share at least one CKC.} \label{Figure - Two type-2 irreducible loops of $F_1$ that share at least one CKC}
\end{figure}
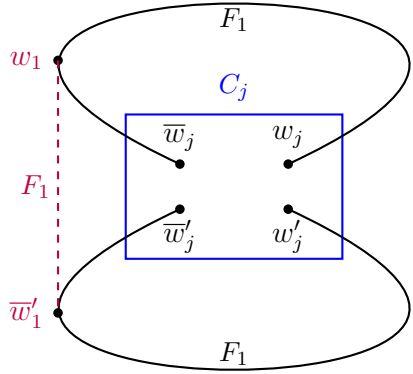

Let us now compose a type-2 irreducible $F_1$ loop, using the fact that $F_1(\o) = F_1(\o')$.
Without loss of generality, assume that $\o=\o_1$ and $\o'=\overline{\o}_1'$, and that $\o_1$ is not in $C_j$.
Moreover, it cannot be the case that $\o_1$ and $\overline{\o}_1'$ are both in the same loop, say $L_1$, because $L_1$ is also an $F_2$-loop and that would imply that either $F_2(\o) = F_2(\o')$ in case $\overline{\o}_1' =\overline{\o}_m$, or that $L_1$ is not a type-2 irreducible loop in case $\overline{\o}_1' \neq \overline{\o}_m$.
Also, it must be that $F_1(\overline{\o}_1') = F_1(\o^*)$ where $\o^*\in L_1$ if and only if $\o^*\in \{\o_1,\overline{\o}_m\}$, otherwise $L_1$ is not type-2 irreducible.

We now split the proof into four possibilities:
\begin{itemize}
    \item $\overline{\o}_1'\in C_j$.
    \item $\overline{\o}_1'\notin C_j$ and $|\{\o_j,\overline{\o}_j\}\cap \{\o_j',\overline{\o}_j'\}|=0,1,2$.
\end{itemize}

Assume that $\overline{\o}_1'\in C_j$.
Consider the loop $(\o_1,\overline{\o}_1,\o_2,\overline{\o}_2,\dots,\o_j,\overline{\o}_1')$.
This loop matches $L_1$ up to state $\o_j$ and $F_1(\o_1)= F_1(\overline{\o}_1')$.
Thus, it is a well-defined type-2 irreducible $F_1$-loop, hence also an $F_2$-loop.
Therefore, $F_2(\o_1)= F_2(\overline{\o}_1')$ and we reach a contradiction.

Moving on to the next possibility, assume that $\overline{\o}_1'\notin C_j$ and $|\{\o_j,\overline{\o}_j\}\cap \{\o_j',\overline{\o}_j'\}|=0$.
Consider the loop $(\o_1,\overline{\o}_1,\o_2,\overline{\o}_2,\dots,\o_j,\overline{\o}_{j}', \o_{j+1}', \overline{\o}_{j+1}', \dots, \o_1',\overline{\o}_1')$.
If  $\o_j$  and $\overline{\o}_{j}'$ are in different partition elements of $F_1$, then this is a well-defined \emph{$F_1$-fully-informative} loop.
If the two states are in the same partition element, then we can omit this pair from the loop and get a shorter loop (in terms of pairs).
This process could be done repeatedly, until we get a well-defined \emph{$F_1$-fully-informative} loop which starts with $\o_1$ and ends with $\overline{\o}_1'$.
If it is a type-2 irreducible $F_1$-loop, then it is also an $F_2$-loop, and $F_2(\o_1)= F_2(\overline{\o}_1')$.
Thus, assume that it is not type-2 irreducible, which implies that it has at least four states in the same partition element of $F_1$.
These four states include neither $\o_1$ nor $\overline{\o}_1'$, because that would imply that either $L_1$ or $L_1'$ is not type-2 irreducible.
Now we can apply Claim \ref{Claim - decomposition of FI loops to type-2 loop}, to decompose this \emph{$F_1$-fully-informative} loop to type-2 irreducible $F_1$-loops, where at least one maintains the connection between $\o_1$ nor $\overline{\o}_1'$ (see the comment at the end of the proof of Claim \ref{Claim - decomposition of FI loops to type-2 loop}).
We thus conclude that it is also an $F_2$-loop and $F_2(\o_1)= F_2(\overline{\o}_1')$.

The next possibility is that $\overline{\o}_1'\notin C_j$ and $|\{\o_j,\overline{\o}_j\}\cap \{\o_j',\overline{\o}_j'\}|=1$.
If either $\o_j' \in \{\o_j,\overline{\o}_j\}$ or $\overline{\o}_j'=\overline{\o}_j$ , then we can follow a similar proof as in the previous case where $|\{\o_j,\overline{\o}_j\}\cap \{\o_j',\overline{\o}_j'\}|=0$, so assume that $\overline{\o}_j' =\o_j$.
In that case, we can redefine the previous loop by omitting $\o_j$ and
$\overline{\o}_j'$ to get $(\o_1,\overline{\o}_1,\o_2,\overline{\o}_2,\dots,\overline{\o}_{j-1}, \o_{j+1}', \overline{\o}_{j+1}', \dots, \o_1',\overline{\o}_1')$.
Again, this is either a well-defined \emph{$F_1$-fully-informative} loop, or could be reduced to such a loop.
Applying the same arguments as before, we conclude that there exists a type-2 irreducible $F_1$-loop which maintains the connection between $\o_1$ nor $\overline{\o}_1'$, so it is also an $F_2$-loop and $F_2(\o_1)= F_2(\overline{\o}_1')$.

The last possibility is that $\overline{\o}_1'\notin C_j$ and $|\{\o_j,\overline{\o}_j\}\cap \{\o_j',\overline{\o}_j'\}|=2$, but in that case the analysis in the previous possibilities holds, and we reach the same conclusion that $F_2(\o_1)= F_2(\overline{\o}_1')$, as needed.\footnote{Note that the proof of Claim \ref{Claim - type-2 irreducible $F_1$-loop with joint CKC} also holds if $\o$ and $\o'$ are not in the original $L_1$ and $L_1'$ loops, respectively, but are simply states in different CKCs that these loops intersect.
That is, if $\o$ and $\o'$ are in different CKCs that $L_1$ and $L_1'$ intersect and $F_1(\o)= F_1(\o')$, we can construct an $F_1$-fully-informative loop that starts with $\o$ and ends with $\o'$ in a similar manner as before, and eventually conclude that $F_2(\o)= F_2(\o')$. \label{footnote - general result for states in connected loops}}
\hfill
\end{proof}

Next, we extend the result of Claim \ref{Claim - type-2 irreducible $F_1$-loop with joint CKC} to more than two loops.
Specifically, we say that two loops $L_i$ and $L_i'$ are \emph{connected} if either they share at least one CKC, or there exists a sequence of loops starting with $L_i$ and ending with $L_i'$ where each two consecutive loops share at least one CKC.

\begin{claim} \label{Claim - transitive closure of loops is F2 mesaurable}
    Consider a set $A$ of type-2 irreducible and connected $F_1$-loops, i.e., every two loops are connected by one of these type-2 irreducible loops.
    Then, every $F_2$-measurable $\tau_2|_{A}$ is $F_1$-measurable.
\end{claim}

\begin{proof}
Let us prove this by induction on the number of loops.
The case of two loops is proved in Claim \ref{Claim - type-2 irreducible $F_1$-loop with joint CKC}, so assume the statement holds for $m$ loops, and consider a set of $m+1$ type-2 irreducible and connected $F_1$-loops.
Further assume, by contradiction, that there exists an $F_2$-measurable strategy over this set that is not $F_1$-measurable.
Thus, there exists $\o$ and $\o'$ such that $F_2(\o)\neq F_2(\o')$ whereas $F_1(\o) = F_1(\o')$.
Evidently, $\o$ and $\o'$ are in different loops and different CKCs.
Denote the loops of $\o$ and $\o'$ by $L_1$ and $L_1'$, respectively.

If $L_1$ and $L_1'$ are connected directly (through a joint CKC) or through at most $m$ loops (including $L_1$ and $L_1'$), then the induction hypothesis holds and every $F_2$-measurable strategy this set of loops is $F_1$-measurable, implying that $F_2(\o) = F_2(\o')$.
Thus, assume that $L_1$ and $L_1'$ are connected through a sequence of all the $m+1$ loops (including $L_1$ and $L_{m+1}$).
Note that $\o'$ cannot be the in the same partition element as any other state from this set of loops, other than $\o$, the state connected to $\o$ in $L_1$, and the state connected to $\o'$ in $L_1'$.
Otherwise, either one of these loops is not type-2 irreducible, or the $F_2$-measurability constraints with every intermediate loop is met (by the induction hypothesis), and again we get that $F_2(\o) = F_2(\o')$.

Thus, we can now follow the same stages as in the proof of Claim \ref{Claim - type-2 irreducible $F_1$-loop with joint CKC} and generate an $F_1$-fully-informative loop based on the sequence of loops connecting $L_1$ and $L_1'$ (as well as $\o$ and $\o'$), which starts with $\o_1$ and ends with $\overline{\o}_1'$.
In this case, Claim \ref{Claim - decomposition of FI loops to type-2 loop} holds and we get a type-2 irreducible $F_1$-loop, which starts with $\o_1$ and ends with $\overline{\o}_1'$, that is also an $F_2$-loop.
We therefore conclude that $F_2(\o) = F_2(\o')$ and the induction follows accordingly.
    \hfill
\end{proof}

After we established that every $F_2$-measurable strategy over a set of connected loops is $F_1$-measurable, let us extend this result to all the CKCs that these loops intersect.
For that purpose, let $A$ be a maximal set of connected loops, where every two are connected, and let $C_A$ be the set of all CKCs that intersect one of these loops (that is, every CKC contains a pair of states from one of these loops).
We refer to every $C_A$ as a \emph{cluster}.
We argue that every $F_2$-measurable strategy over a cluster $C_A$ is $F_1$-measurable.
To see this, recall Footnote \ref{footnote - general result for states in connected loops} which states that the proof of Claim \ref{Claim - type-2 irreducible $F_1$-loop with joint CKC}  holds for every $\o$ and $\o'$ in two different CKCs that intersect two connected loops $L_1$ and $L_1'$, respectively.
Namely, for every two such states $\o$ and $\o'$ where $F_1(\o) = F_1(\o')$, it follows that $F_2(\o) = F_2(\o')$.
So, as argued in the proof of Claim \ref{Claim - transitive closure of loops is F2 mesaurable}, we conclude that every $F_2$-measurable strategy over a cluster is $F_1$-measurable.

\begin{obs} \label{Observation - every tau2 over CA is F1-measurable.}
    Every $F_2$-measurable strategy over a cluster is $F_1$-measurable.
\end{obs}

Once we have established that every $F_2$-measurable strategy over a cluster is $F_1$-measurable, let us consider a partition $\Omega^*$ of $\Omega$ into clusters and individual CKCs that are not part of clusters.
Note that \emph{any} two elements of the partition $\Omega^*$ jointly intersect at most one partition element of $F_1$, otherwise the two components would be in the same cluster.
To see this, consider the different possible intersections of elements in $\Omega^*$.
If both elements $A_1$ and $A_2$ are CKCs, then any two different partition elements of $F_1$ that intersect both $A_1$ and $A_2$ would form a type-2 irreducible $F_1$-loop.
Otherwise, one of these elements is a cluster, say $A_1$, and it follows from previous proofs that for every $\o$ and $\o'$ that belong to the same cluster (but in different CKCs) and $F_1(\o)=F_1(\o')$, then one can form an $F_1$-fully-informative loop that starts with $\o$ and ends with $\o'$.
Thus, in case $\o $ and $\o'$ are in cluster $A_1$ and in different partition elements of $F_1$ that intersect $A_2$ (whether $A_2$ is a CKC or another cluster), one can form an $F_1$-fully-informative loop that intersects $A_1$ and $A_2$.
Using Claim \ref{Claim - decomposition of FI loops to type-2 loop}, we can conclude that $A_1$ and $A_2$ belong to the same cluster.
This result is summarized in the following observation.

\begin{obs} \label{Observation - single connecting element between clusters}
    Fix two elements $A_1,A_2 \in \Omega^*$.
    Then, there exists at most one partition element $F_1(\o)$ of $F_1$ such that $F_1(\o)\cap A_1$ and $F_1(\o)\cap A_2$ are non-empty sets.
\end{obs}

We would now want to prove that Oracle $1$ can mimic every $F_2$-measurable strategy defined over $\Omega^*$.
For this purpose, we present the following Lemma \ref{Lemma - cross-section lemma} which relates to the $F_2$-measurability constraints over different sets of CKCs, that are not in the same cluster (i.e., they are not connected by type-2 irreducible $F_1$-loops).

\begin{lemma} \label{Lemma - cross-section lemma}
    Fix two disjoint sets $A_1,A_2 \subseteq \Omega$ that do not intersect the same \emph{CKCs}, and denote $A= A_1\cup A_2$.
    Assume that:
    \begin{itemize}
        \item For every $i$ and for every $F_2$-measurable $\tau_2|_{A_i}$, there exists an $F_1$-measurable $\tau_1^i|_{A_i}$, such that $\mu_{\tau_1}|_{A_i}=\mu_{\tau_2}|_{A_i}$.
        \item For every $\o_{1},\o_{1}' \in A_1$ and $\o_{2},\o_{2}' \in A_2$ such that $F_1(\o_1)=F_1(\o_2)$ and $F_1(\o_1')=F_1(\o_2')$, it follows that $F_1(\o_1)=F_1(\o_1')$.
    \end{itemize}
    Then, for every $\tau_2|_{A}$, there exists $\tau_1|_{A}$ such that $\mu_{\tau_1}|_{A_i}=\mu_{\tau_2}|_{A_i}$ for every $i=1,2$.
\end{lemma}

\begin{proof}
Fix $\tau_2|_A$ and $\tau_1^i|_{A_i}$ where $i=1,2$, such that $\mu_{\tau_2}|_{A_i} = \mu_{\tau_1^i}|_{A_i}$  for every $i$.
Define the sets $\tilde{A}_i=\{\o_i \in A_i: \exists \o_{-i} \in A_{-i},  F_1(\o_i)=F_1(\o_{-i}) \}$ for every $i=1,2$.
The second condition of the claim implies that all the states in $\tilde{A}_1\cup \tilde{A}_2$ are in the same partition element of $F_1$.
To see this, fix $\o_1 \in \tilde{A}_1$ and, by definition, there exists a state $\o_2 \in \tilde{A}_2$ such that $F_1(\o_1)=F_2(\o_2)$.
If there exists another $\o_1' \in \tilde{A}_1$, it is either connected to $\o_2$ (i.e., $F_1(\o_1')=F_1(\o_2)$), or to some $\o_2' \in \tilde{A}_2$, and in that case the condition implies that $F_1(\o_1)=F_1(\o_1')$.
The same holds for every $\o_2 \in \tilde{A}_2$

For every $i=1,2$, let $S_i$ be the signals induced by $\tau_1^i|_{A_i}$.
Define the following strategy $\tau_1$:
    \begin{eqnarray*}
        \tau_1((s_1,s_2)|\o)  =
        \begin{cases}
            \tau_1^1(s_1|\o)\tau_1^2(s_2|\tilde{A}_2), & \text{if \ $\o \in A_1, (s_1,s_2) \in S_1\times  S_2 $},\\
		\tau_1^1(s_1|\tilde{A}_1) \tau_1^2(s_2|\o), & \text{if \ $\o \in A_2, (s_1,s_2) \in S_1\times  S_2 $}.
	\end{cases}
    \end{eqnarray*}
One can easily verify that $\sum_{(s_1,s_2)} \tau_1((s_1,s_2)|\o)=1$ for every $\o$, so $\tau_1$ is indeed a strategy.

Let us now prove that $\tau_1$ is $F_1$-measurable and $\mu_{\tau_1}|_{A}=\mu_{\tau_2}|_{A}$.
If we restrict $\tau_1$ to $A_i$, it is clearly $F_1$-measurable as $\tau_1^{-i}(s_{-i}|\tilde{A}_{-i})$ is fixed for every $\o \in A_i$ and $s_i\in S_i$.
Thus, consider $\tau_1((s_1,s_2)|\o)$ where $\o\in \tilde{A}_1$.
All the states in $\tilde{A}_1\cup \tilde{A}_2$ are in the same partition element of $F_1$, so for every $(\o_1,\o_2)\in \tilde{A}_1\times \tilde{A}_2$ we get
\begin{eqnarray*}
    \tau_1((s_1,s_2)|\o_1)
    & = & \tau_1^1(s_1|\o_1)\tau_1^2(s_2|\tilde{A}_2) \\
    & = & \tau_1^1(s_1|\tilde{A}_1)\tau_1^2(s_2|\tilde{A}_2) \\
    & = & \tau_1^1(s_1|\tilde{A}_1)\tau_1^2(s_2|\o_2) \\
    & = & \tau_1((s_1,s_2)|\o_2),
\end{eqnarray*}
and the $F_1$-measurability condition holds.
Moreover, for every $\o_i,\o_i'\in A_i$ and for every $(s_1,s_2)$ such that $\tau_1^i(s_i|\o)>0$ where $\o\in \{\o_1,\o_1'\}$, it follows that
$$
\frac{\tau_1((s_1,s_2)|\o_i,A_i)}{\tau_1((s_1,s_2)|\o_i',A_i)} =\frac{\tau_1^i(s_i|\o_i)}{\tau_1^i(s_i|\o_i')},
$$
which implies that conditional on $A_i$, $\tau_1$ yields the same distribution over posteriors profiles as $\tau_1^i$, thus mimicking $\tau_2$ on every $A_i$, as needed.
\hfill
\end{proof}

We can thus finalize the proof using induction on the number of elements in $\Omega^*$.
Until now, we established in Observation \ref{Observation - every tau2 over CA is F1-measurable.}, Observation \ref{Observation - single connecting element between clusters} and Lemma \ref{Lemma - cross-section lemma} that, given either $|\Omega^*|=1$ or $|\Omega^*|=2$, then for every $F_2$-measurable strategy $\tau_2|_{\Omega^*}$, there exists  $\tau_1|_{\Omega^*}$ such that $\mu_{\tau_1}|_{A}=\mu_{\tau_2}|_{A}$ for every $A\in \Omega^*$.
Assume this holds for $|\Omega^*|=k \geq 2$, and consider $|\Omega^*|=k+1$.

Denote the elements of $\Omega^*$ by $A_1,A_2,\dots,A_k,A_{k+1}$.
If there exists only one partition element of $F_1$ that intersects $A_{k+1}$ and at least one $A_i$ for $i\leq k$, then Lemma \ref{Lemma - cross-section lemma} holds and the result follows.
Thus, assume there are at least two different partition elements $F_1(\o)=F_1(\o_1)$ and $F_1(\o')=F_1(\o_2)$ of $F_1$ such that $\o,\o' \in A_{k+1}$ and $\o_i \in A_i $ for every $i=1,2$.

The proof now splits into two parts: either $A_1$ and $A_2$ are connected (i.e., there exists a sequence of partition elements of $F_1$ that sequentially intersect elements in $\Omega^*\setminus A_{k+1}$, starting with $A_1$ and ending with $A_2$) or $A_1$ and $A_2$ are unconnected.
If they are unconnected, we can apply Lemma \ref{Lemma - cross-section lemma} for $A_{1}$ and $A_{k+1}$ and then use the induction hypothesis, so we assume they are connected.

Whether $A_{k+1}$ is a CKC or a cluster and assuming that $A_1$ and $A_2$ are connected, we argue that there exists a type-2 irreducible $F_1$-loop that include $\o$ and $\o'$, implying that $A_{k+1}$ is part of a cluster with other elements in $\Omega^*$.
To see this, recall whenever $\o$ and $\o'$ belong to the same cluster and $F_1(\o)=F_1(\o')$, then there exists an $F_1$-fully-informative loop that start with $\o$ and ends with $\o'$.
So consider such a sequence of states $l_{\o \to \o'} = (\o,\dots,\o')$, which would have been an $F_1$-loop had $F_1(\o)=F_1(\o')$.

Next, fix the entire path of connections of elements in $\Omega^*$ that starts with $A_1$ and ends with $A_2$.
Again, the connection between $A_1$ and $A_2$ implies that there exists a sequence of states $l_{\o_1\to \o_2}=(\o_1,\dots,\o_2)$ in $\Omega^*\setminus A_{k+1}$, that would have been an $F_1$-loop had $F_1(\o_1)=F_1(\o_2)$.
Hence, consider the sequence of states $l=(\o,\dots,\o',\o_2\dots,\o_1)$ which forms an informative $F_1$-loop, because $F_1(\o) \neq F_1(\o')$.
Using Proposition \ref{Proposition - irreducible and informative loops} and Claim \ref{Claim - decomposition of FI loops to type-2 loop}, we know that this loop has a type-2 irreducible $F_1$-sub-loop that contains $\o$ and $\o'$.
Thus, $A_{k+1}$ is in the same cluster as other elements in $\Omega^*$, thus contradicting the assumption that $|\Omega^*|=k+1$.
\hfill
\end{proof}

\end{document}